\documentclass{ws-ijmpcs}

\begin{document}

\markboth{R. Mankel}
{Higgs Searches Beyond the Standard Model}

%
\catchline{}{}{}{}{}
%

\title{Higgs Searches Beyond the Standard Model}

\author{R.~Mankel, on behalf of the ATLAS, CMS, CDF and D0 collaborations}

\address{Deutsches Elektronen-Synchrotron (DESY),\\ 
Notkestr 85, D-22603 Hamburg, Germany\\
Rainer.Mankel@desy.de}

\maketitle

\begin{history}
\received{Day Month Year}
\revised{Day Month Year}
\end{history}

\begin{abstract}
  While the existence of a Higgs boson with a mass near 125~GeV has
  been clearly established, the detailed structure of the entire Higgs
  sector is yet unclear. Besides the Standard Model interpretation,
  various possibilities for extended Higgs sectors are being
  considered. The minimal supersymmetric extension (MSSM) features two
  Higgs doublets resulting in five physical Higgs bosons, which are
  subject to direct searches. Alternatively, more generic Two-Higgs Doublet
  models (2HDM) are used for the interpretation of results. The
  Next-to-Minimal Supersymmetric Model (NMSSM) has a more complex
  Higgs sector with seven physical states. Also exotic Higgs bosons
  decaying to invisible final states are considered. This article
  summarizes recent findings based on results from collider experiments.
\end{abstract}

\ccode{PACS numbers:}

\section{Introduction}	
\begin{figure}[htbp]
\centerline{\includegraphics[width=6cm]{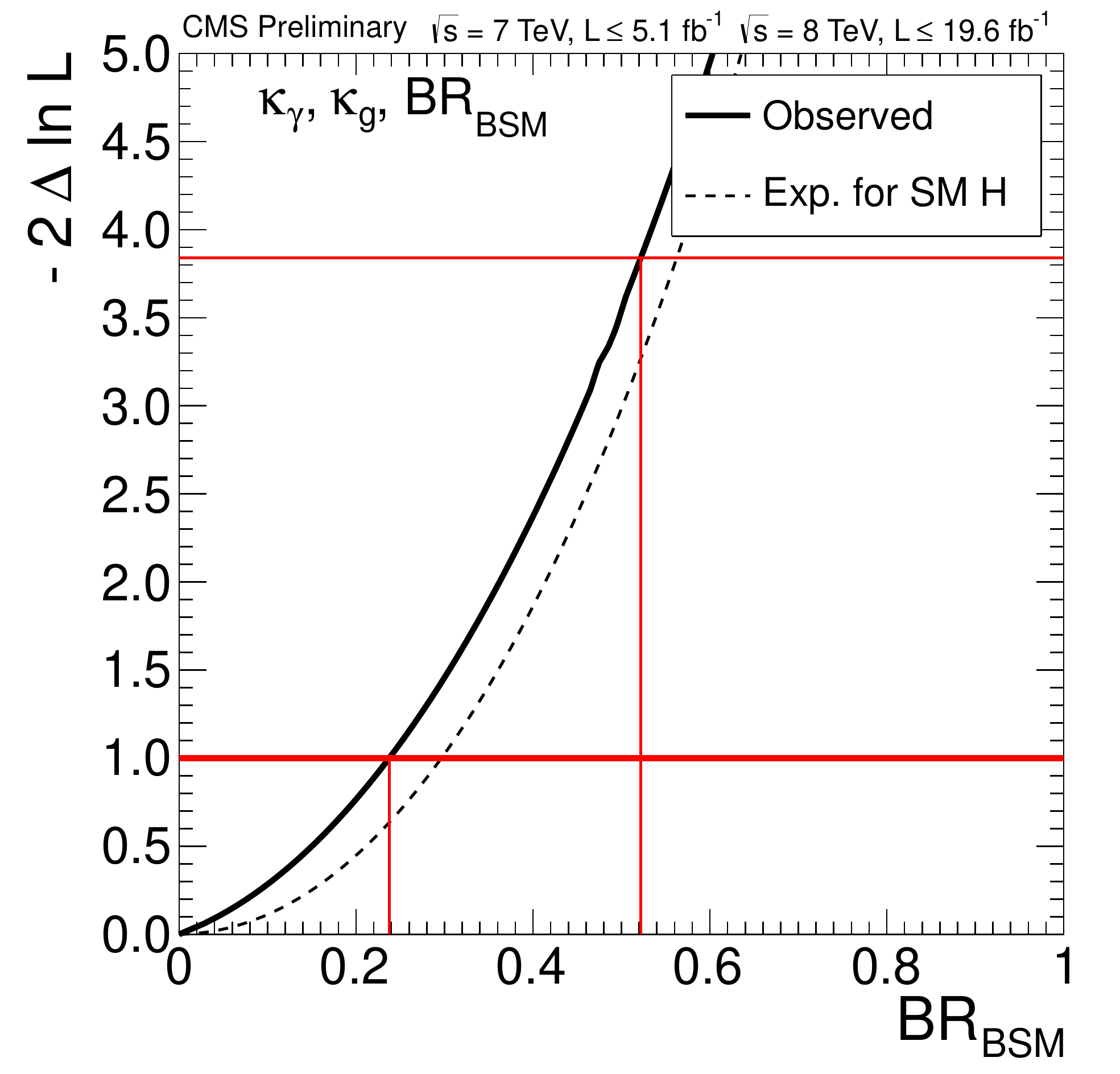}}
\vspace*{8pt}
\caption{1D test statistics as a function of the Higgs branching
  fraction into non-SM decay modes\protect\cite{CMS_BR_BSM}. Only the photon and gluon
  couplings within the loops are modified. The tree-level couplings
  are fixed at SM level.}
\label{Fig:BR_BSM}
\end{figure}

The electroweak symmetry breaking mechanism of the Standard Model (SM)
predicts the Higgs particle as a scalar boson. The discovery of a
Higgs boson with a mass near
125~GeV\cite{HiggsDiscoveryATLAS,HiggsDiscoveryCMS} by the 
experiments ATLAS\cite{ATLASJinst} and CMS\cite{CMSJinst}  at the CERN Large
Hadron Collider (LHC) is an important milestone, and it is of
fundamental interest to study the properties of this state, its
quantum numbers and couplings. An equally important aim is the
unravelling of the overall structure of the Higgs sector. While at the
level of current measurements, the observed state is compatible with
the Higgs boson as predicted by the SM, the mass of
the Higgs boson is divergent at high energies\cite{massDivergence}.
For many other open
questions, related e.g. to dark matter, naturalness and CP violation
in the universe, solutions might evolve through more detailed studies
of the Higgs sector. From investigation of the observed boson alone
one obtains relatively weak contraints on Higgs decays beyond the
Standard Model (BSM); for example, a study of the CMS experiment
(Figure~\ref{Fig:BR_BSM})\cite{CMS_BR_BSM} derives an upper limit of
52\% for the branching fraction into BSM decay modes
at 95\% confidence level. Significant improvement of such limits can at least be
expected to take a long time. For this reason, direct searches of BSM
Higgs signatures are essential to clarify the structure of the Higgs
sector.

\section{The MSSM Higgs Sector}
While the SM features a single complex Higgs doublet
resulting in one physical scalar Higgs boson, the Minimal
Supersymmetric Model (MSSM) is characterized by an extended Higgs
sector with two complex Higgs doublets. This gives rise to five 
physical Higgs bosons, three of them neutral, labeled h, H (CP-even) and A
(CP-odd) and jointly referred to as $\phi$, and the other two
charged, named H$^\pm$. At tree level, the MSSM Higgs sector is governed by
two parameters: the mass $m_A$ and $\tan \beta$, which is the ratio of
the vacuum expectation values of the two Higgs doublets. Beyond tree
level, additional parameters enter via radiative corrections, and
benchmark scenarios fixing these parameters are used to compare
different measurements. In many cases, the $m_h^{max}$ benchmark
scenario\cite{mhmax} is used.

The mass of the CP-odd Higgs boson is usually approximately degenerate
with one of the CP-even bosons, either the H for large $m_A$, or the h
if $m_A$ is small. With the exception of the di-muon channel, this
degeneracy cannot be resolved within the experimental mass resolution,
and thus the visible cross section effectively doubles. The coupling
to the b quark is proportional to $\tan \beta$. Thus production cross
sections in association with b quarks are enhanced by a factor of
$\approx 2 \tan^2 \beta$. It is important to note that the observation
of a Higgs boson near 125~GeV with SM-like properties does
not exclude additional heavy Higgs bosons at large $\tan
\beta$, as discussed in detail in Ref.~\refcite{CarenaMhMod}. If $m_A$
is large compared to the Z mass,
the light MSSM Higgs boson (h) becomes SM-like, a situation
referred to as the decoupling limit. At the current level of measurements,
both SM and MSSM are found to fit the measurements about equally well, see for example Ref.~\refcite{Bechtle}.

\subsection{MSSM searches in the $\tau \tau$ channel}
The $\phi \rightarrow \tau \tau$ channel can be seen as a good
compromise between a relatively large branching fraction and
manageable backgrounds. Searches in this channel have been performed by the 
Tevatron\cite{TevatronTauTau} and LHC experiments; the latter will be described in more detail. 
The main production mechanisms are associated
production with b quarks and gluon-gluon fusion (GGF). In correspondence,
the data are sub-divided in event categories with at least one, or no
b-tagged jet. The decay modes of the $\tau$ are grouped into leptonic
($e$ or $\mu$) and hadronic decays (``had''). This results in six decay
patterns, five of which are covered by the analyses of ATLAS\cite{ATLASTauTau} and CMS\cite{CMSTauTau}: $e+\mu$,
$e+had$, $\mu+had$, $had+had$ (only ATLAS), $\mu+\mu$ (only CMS). The
invariant mass of the $\tau$ pair is reconstructed from the visible
decay products and the missing transverse energy, using a likelihood
technique (CMS) and a method named Missing Mass Calculator\cite{MMC}
(ATLAS).

\begin{figure}[htbp]
\centerline{\includegraphics[width=6.0cm]{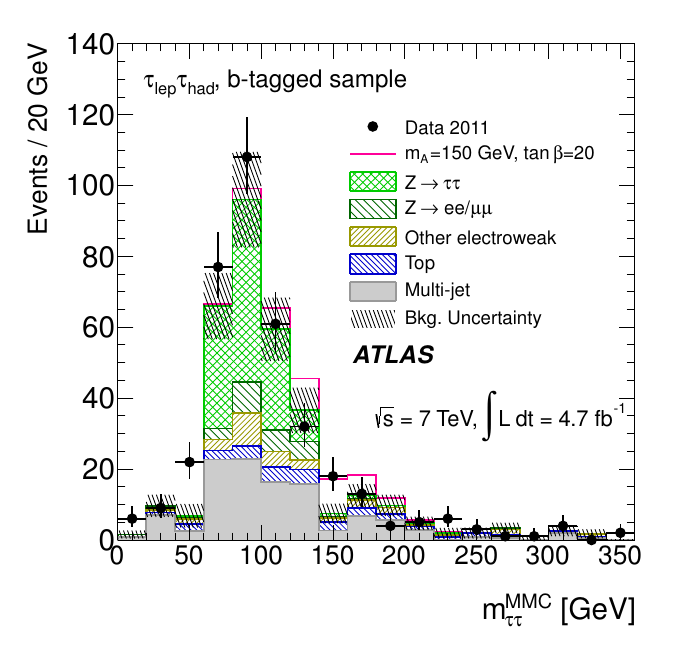}
\includegraphics[width=6.0cm]{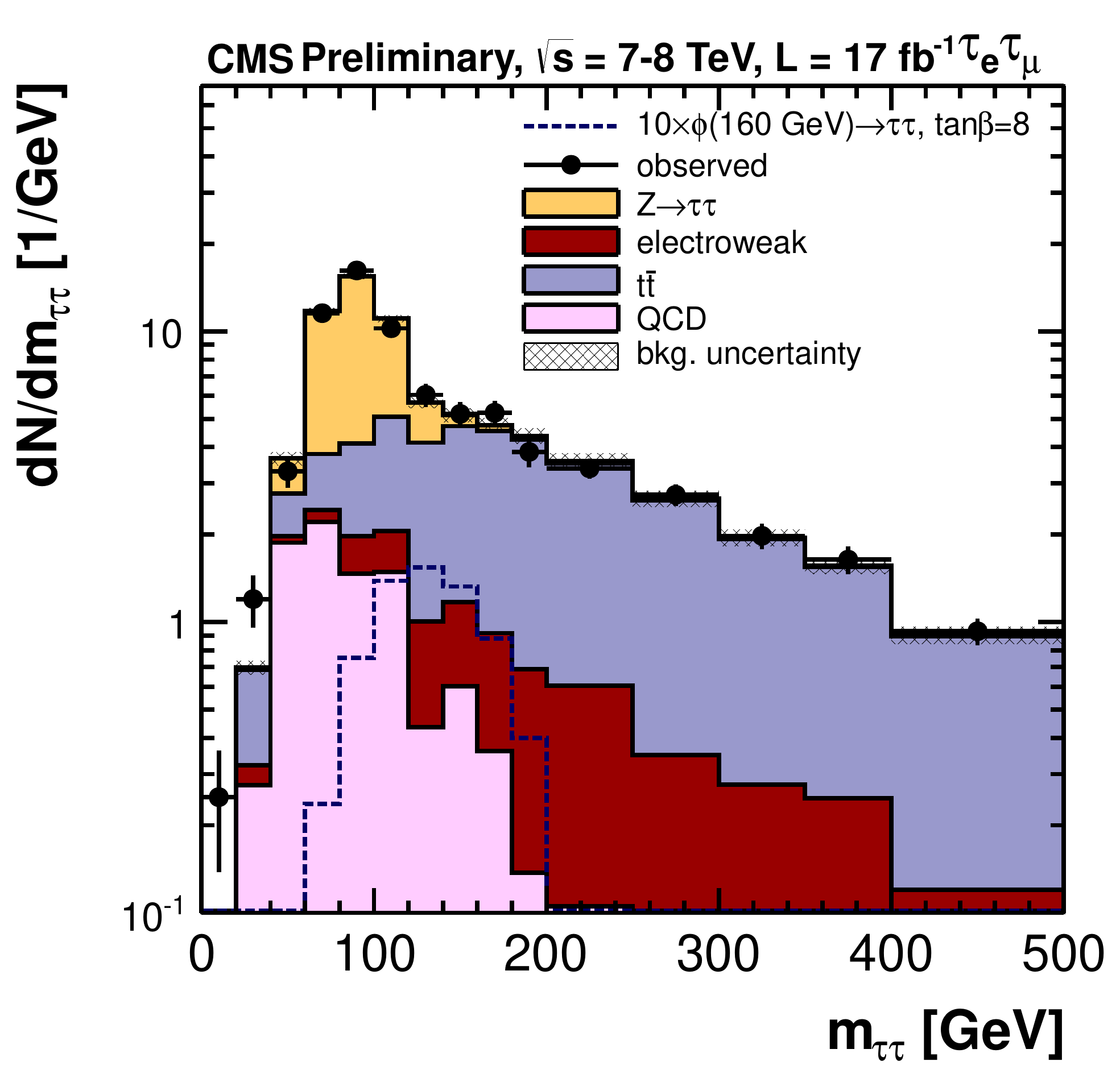}}
\vspace*{8pt}
\caption{Distributions of reconstructed invariant mass of the $\tau
  \tau$ system in the btag category; from
  ATLAS\protect\cite{ATLASTauTau} in the combined $e+had$ and $\mu +
  had$ channels (left) and CMS\protect\cite{CMSTauTau} (right) in the
  $e+\mu$ channel. Estimated backgrounds are also shown.}
\label{Fig:mtautau}
\end{figure}

A very important background arises from $Z \rightarrow \tau \tau$
decays. It is addressed with an embedding technique: events with
reconstructed $Z \rightarrow \mu \mu$ decays are taken from the data,
and the muon objects are replaced with simulated $\tau$
decays. Additional backgrounds arise from $Z \rightarrow ee (\mu \mu)$
decays, $t\bar t$ and di-boson production, QCD multijet and W+jets
events. Their relevance differs across the various $\tau \tau$ decay patterns.
The results shown here are based on the 7~TeV data sample in the case
of ATLAS, while the CMS results include the first part of the 8~TeV
data.
Figure~\ref{Fig:mtautau} shows examples of the reconstructed $m_{\tau
  \tau}$ spectra for selected decay patterns in the b-tag category for
ATLAS and CMS together with the estimated backgrounds. The summed
background estimates describe the data very well, and there is no
indication of a signal.

\begin{figure}[htbp]
\centerline{\includegraphics[width=6.0cm]{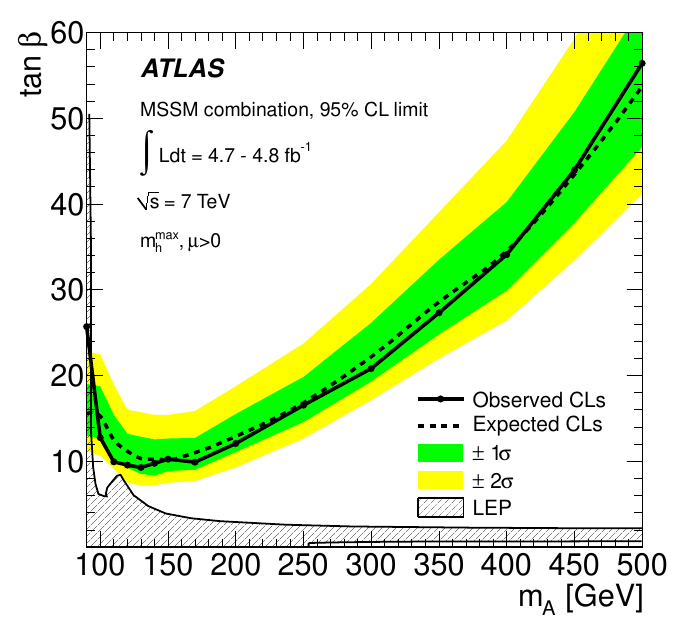}
\includegraphics[width=6.0cm]{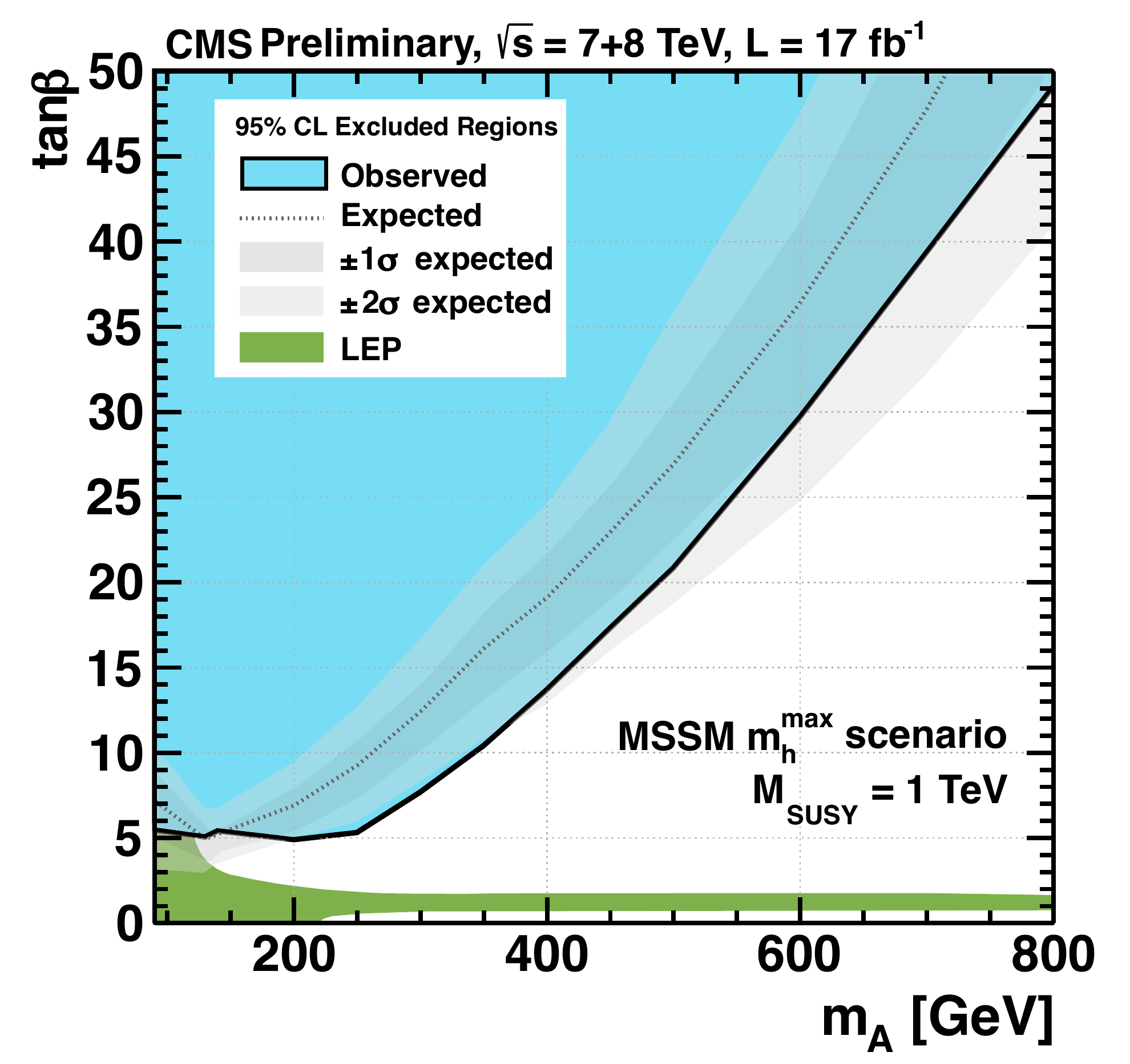}}
\vspace*{8pt}
\caption{Upper limits for $\tan \beta$ vs. $m_A$ obtained in the $\tau \tau$ channel 
  from the ATLAS\protect\cite{ATLASTauTau} (left) and CMS experiments\protect\cite{CMSTauTau} (right). 
  The ATLAS measurement also includes the $\phi \rightarrow \mu \mu$ channel.}
\label{Fig:UL_tautau}
\end{figure}
The results from the various event categories and decay patterns are
combined, and are used to compute upper limits on the MSSM parameter
$\tan \beta$ at the 95\% confidence level as a function of $m_A$. The ATLAS analysis also
includes the $\phi \rightarrow \mu^+\mu^-$ decay channel, which is
not discussed here in detail. The results are shown in
Figure~\ref{Fig:UL_tautau} in the $m_h^{max}$ scenario. At low masses (below $m_A \approx
250$~GeV), the upper limit on $\tan \beta$ reaches down to values of 5, and touches
the lower limit obtained by the LEP experiments. At larger $m_A$,
there is still a wide range of $\tan \beta$ allowed. Inclusion of the
full set of 8~TeV data can be expected to give further improvements in the
sensitivity of both experiments.

\subsection{MSSM searches in the $b\bar b$ channel}
In the MSSM, the Higgs decay into two b quarks is the dominating
channel for $\tan \beta$ significantly larger than one. Furthermore,
the associated Higgs production with b quarks is enhanced by a factor
of $\tan^2 \beta$ due to the modified coupling. On the other hand,
there is a copious background from QCD multi-jet production, which
makes this analysis very difficult and in particular challenging from
the trigger aspect.

\begin{figure}[htbp]
\centerline{\includegraphics[height=6.0cm]{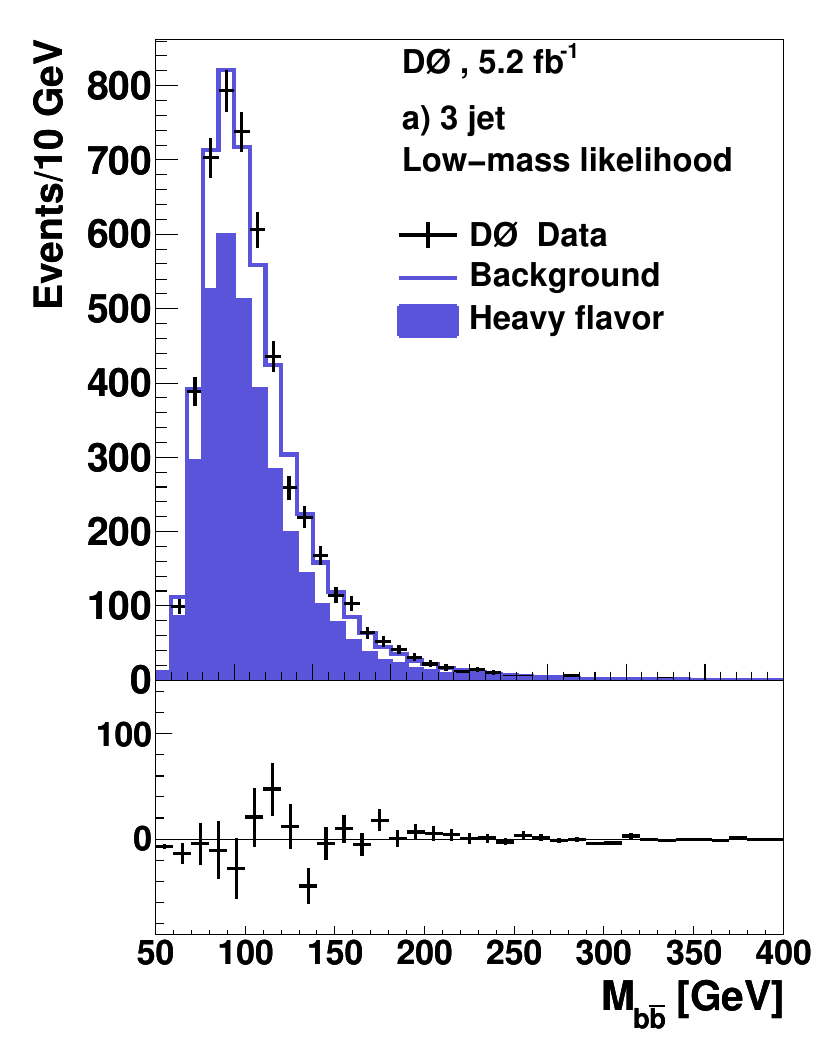}
\includegraphics[height=6.0cm]{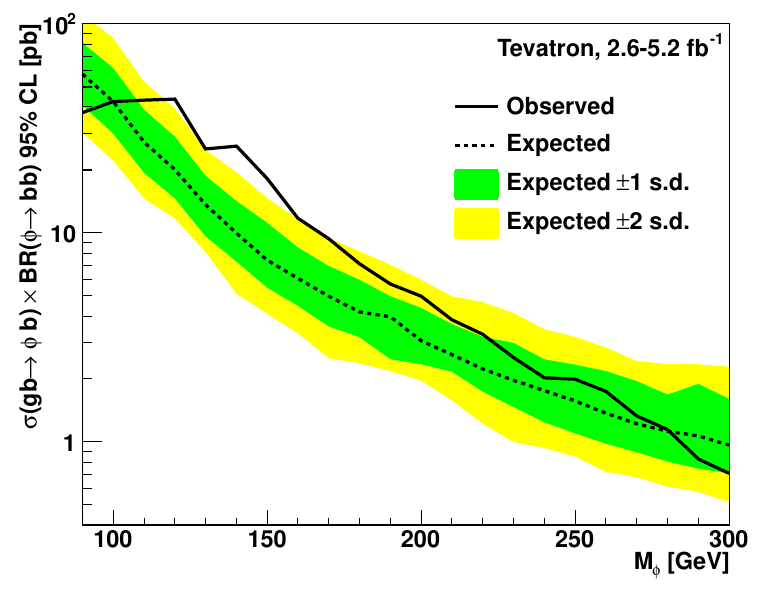}}
\vspace*{8pt}
\caption{Left: Invariant mass spectrum of the two leading b jets in
  triple-b-tag events from the D0 experiment, together with the
  estimated background\protect\cite{bbDZero}. A likelihood-based
  selection optimized for the low-mass region is shown. Right:
  Combined cross section times branching fraction upper limit from the
  CDF and D0 experiments\protect\cite{bbTevatronCombination}.}
\label{Fig:bbTevatron}
\end{figure}

The analyses shown here search for $\phi \rightarrow b\bar b$
signatures with at least one additional b-tagged jet. This channel was first
successfully analyzed by the Tevatron
experiments CDF and D0\cite{CDF_bb,bbDZero,bbTevatronCombination}. The signal is searched in the invariant mass
of the two leading jets, which must be b-tagged. CDF uses a second
variable which functions as a global b-tag. The background estimation, which is a key
component of the analyses, is performed with different methods. The
CDF analysis derives background templates from the double b-tag sample
by applying b-tag efficiency weights. The combination of the various
background templates, together with a signal template, is fitted to
the data. In the D0 analysis, fractional contributions of the various
multi-jet processes are determined by fitting $p_T$ distributions from
simulation to the data. The invariant mass spectrum from the D0
analysis is shown in Figure~\ref{Fig:bbTevatron}~(left)\cite{bbDZero}. The data are
shown along with the data-driven background estimate. The combined
upper limits on the cross section times branching fraction is shown in
Figure~\ref{Fig:bbTevatron}~(right)\cite{bbTevatronCombination}.
Neither experiment sees a signal above the background expectation, but
there are modest excesses in the observed vs. the expected upper
limits of $\approx 2.8\sigma$ in the CDF and of $\approx 2.5\sigma$ in
the D0 case, which are both visible in the combination plot at relatively low masses.

\begin{figure}[htbp]
\centerline{\includegraphics[height=6.0cm]{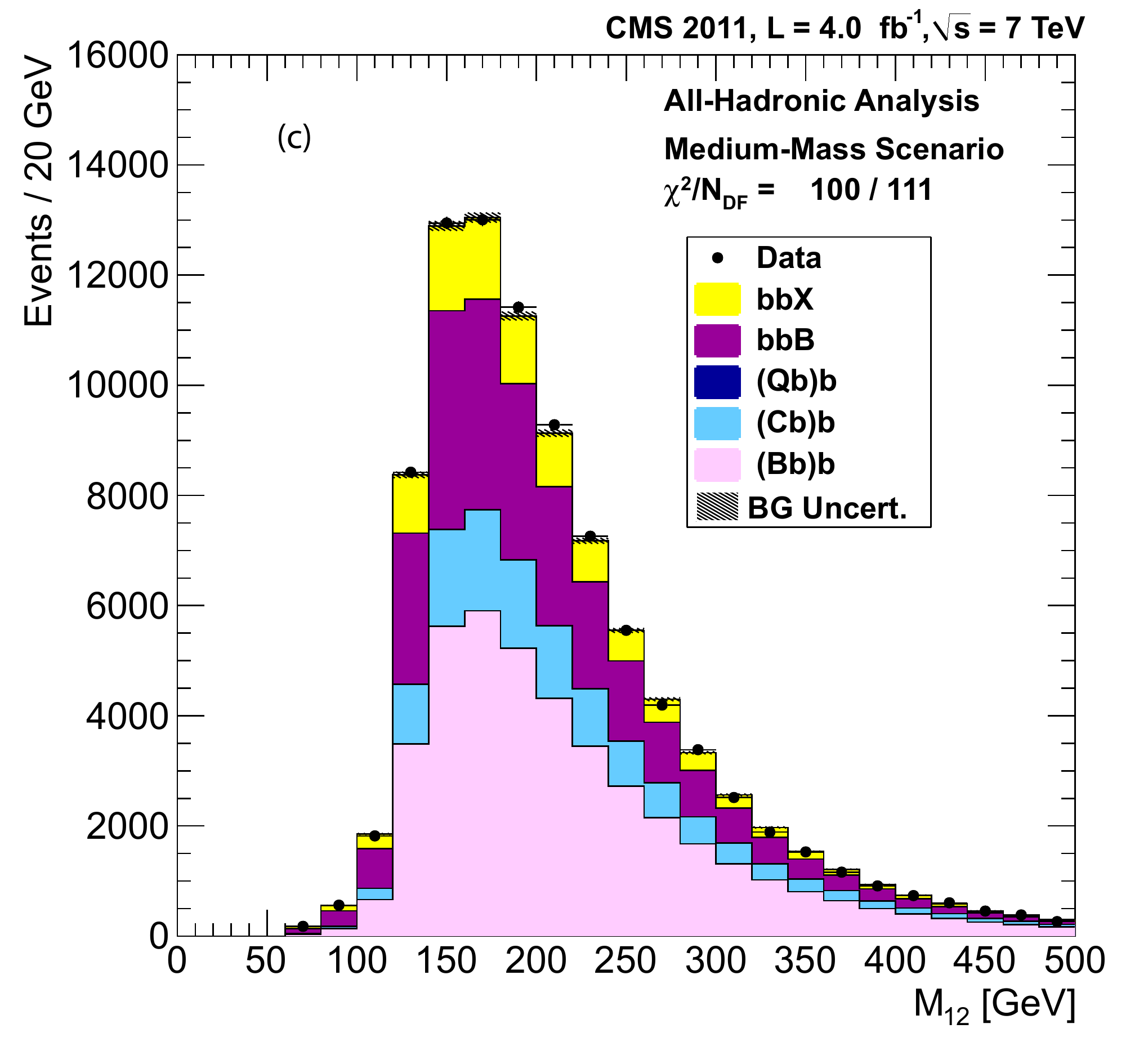}
\includegraphics[height=6.0cm]{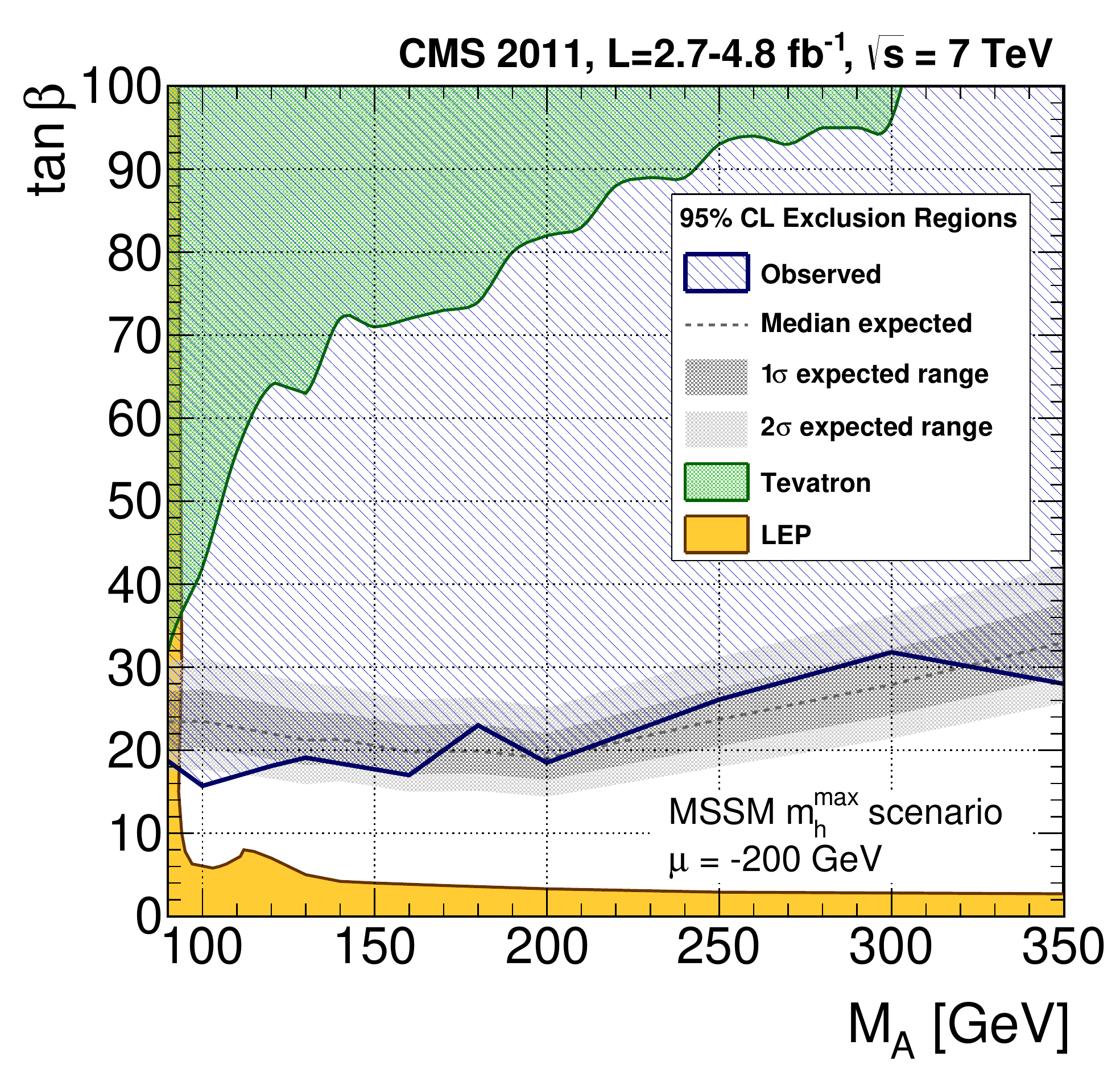}}
\vspace*{8pt}
\caption{Left: Invariant mass spectrum of the two leading b jets in
  triple-b-tag events from the all-hadronic analysis in CMS\protect\cite{CMS_MSSM_Hbb}. The
  shaded stacked histograms show the contribution of the different
  background templates resulting from the background-only fit.
  Right:
  Observed upper limits at 95~\% confidence level on $\tan \beta$ as a
  function of $m_A$ for the combined all-hadronic and semi-leptonic
  analyses from CMS\protect\cite{CMS_MSSM_Hbb}. Exclusion regions from LEP and Tevatron are also shown.}
\label{Fig:CMSHbb}
\end{figure}
The first analysis of this channel at the LHC has been carried out by
CMS\cite{CMS_MSSM_Hbb}. The search is performed both in the
all-hadronic final state with three b-tagged jets, as well as in
semi-leptonic signatures requiring in addition a non-isolated muon in one of the
b-tagged jets. The all-hadronic analysis is inspired by the CDF
method, using a background model of templates determined from the
double-b-tag sample. The invariant mass spectrum of the two leading
jets is shown in Figure~\ref{Fig:CMSHbb} (left). The observed spectrum
can be fitted well with a combination of five background templates,
and there is no indication of a signal or excess. Upper limits for the cross
section times branching ratio are converted to limits in the MSSM
parameter space in the $m_h^{max}$ scenario. In Figure~
\ref{Fig:CMSHbb} (right) these limits are shown as a function of $\tan
\beta$ in comparison to the combined Tevatron results. The CMS numbers
have been converted for a Higgsino mass parameter of $\mu=-200$~GeV to
allow direct comparison with the Tevatron numbers. The CMS results
shows a much higher sensitivity already with the 7~TeV data, with
limits ranging between 18 and 32 in the whole mass range up to
350~GeV.

\subsection{Charged Higgs boson search}

\begin{figure}[htbp]
\centerline{\includegraphics[height=5.0cm]{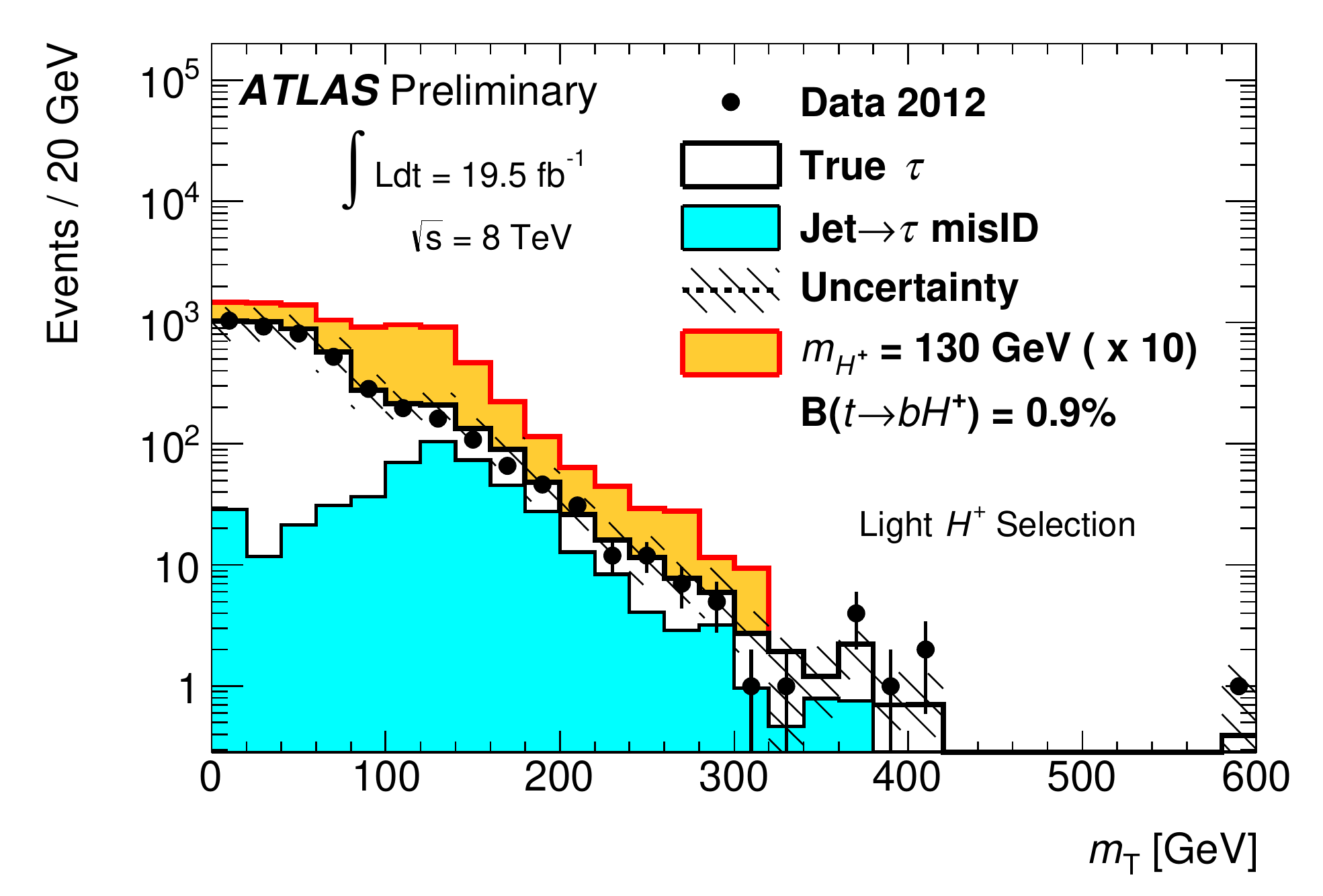}
\includegraphics[height=5.0cm]{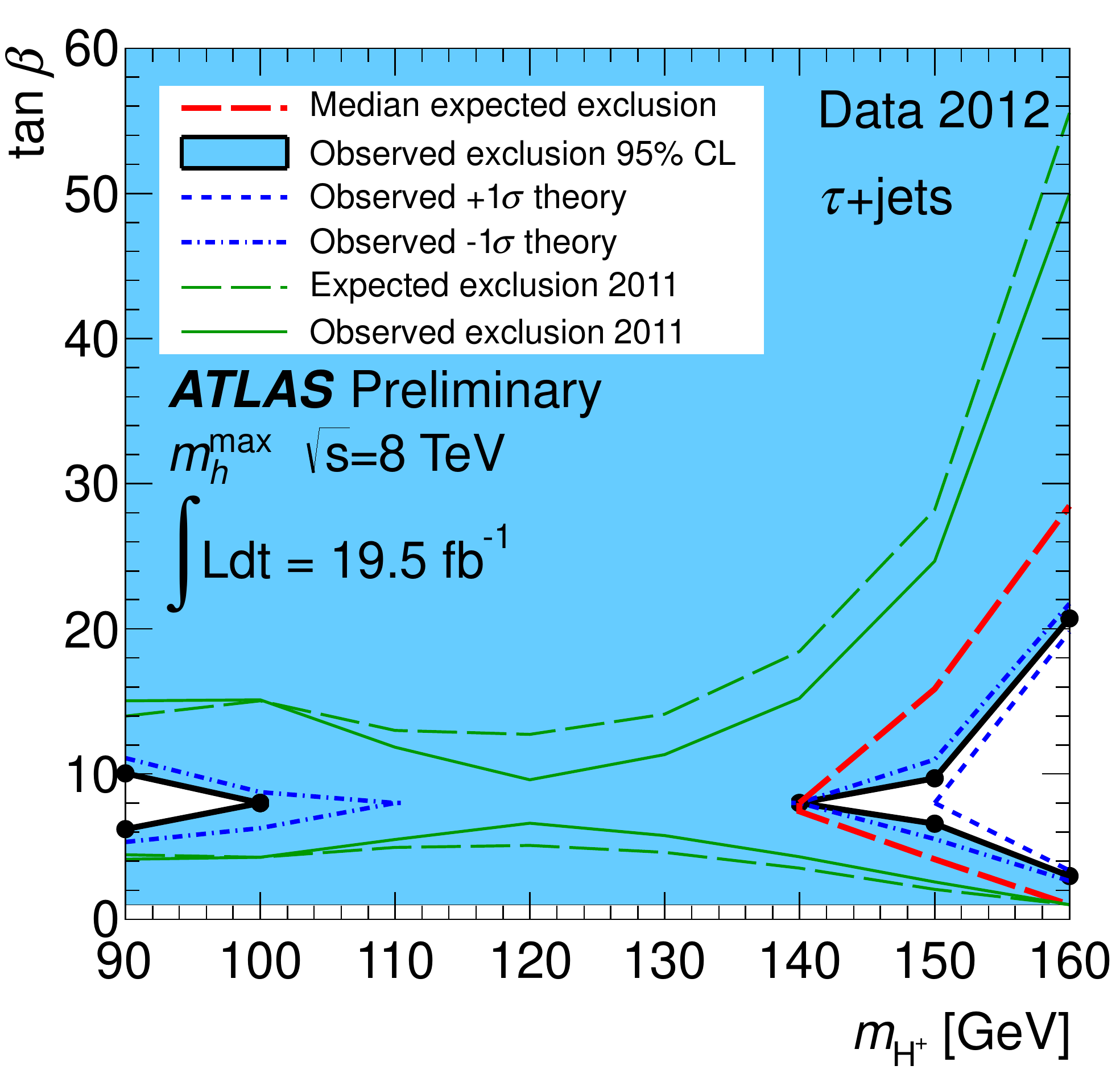}}
\centerline{\includegraphics[height=5.0cm]{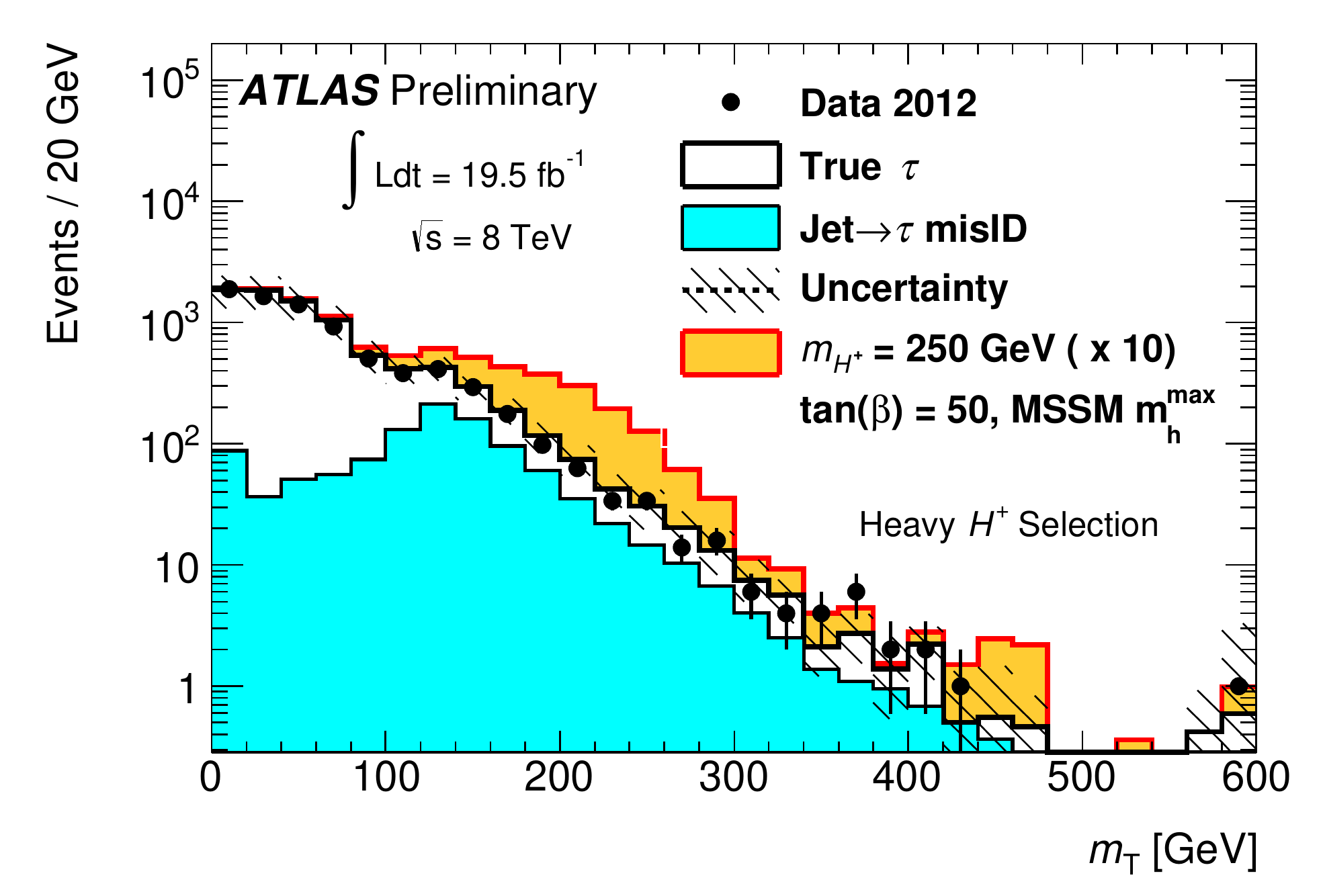}
\includegraphics[height=5.0cm]{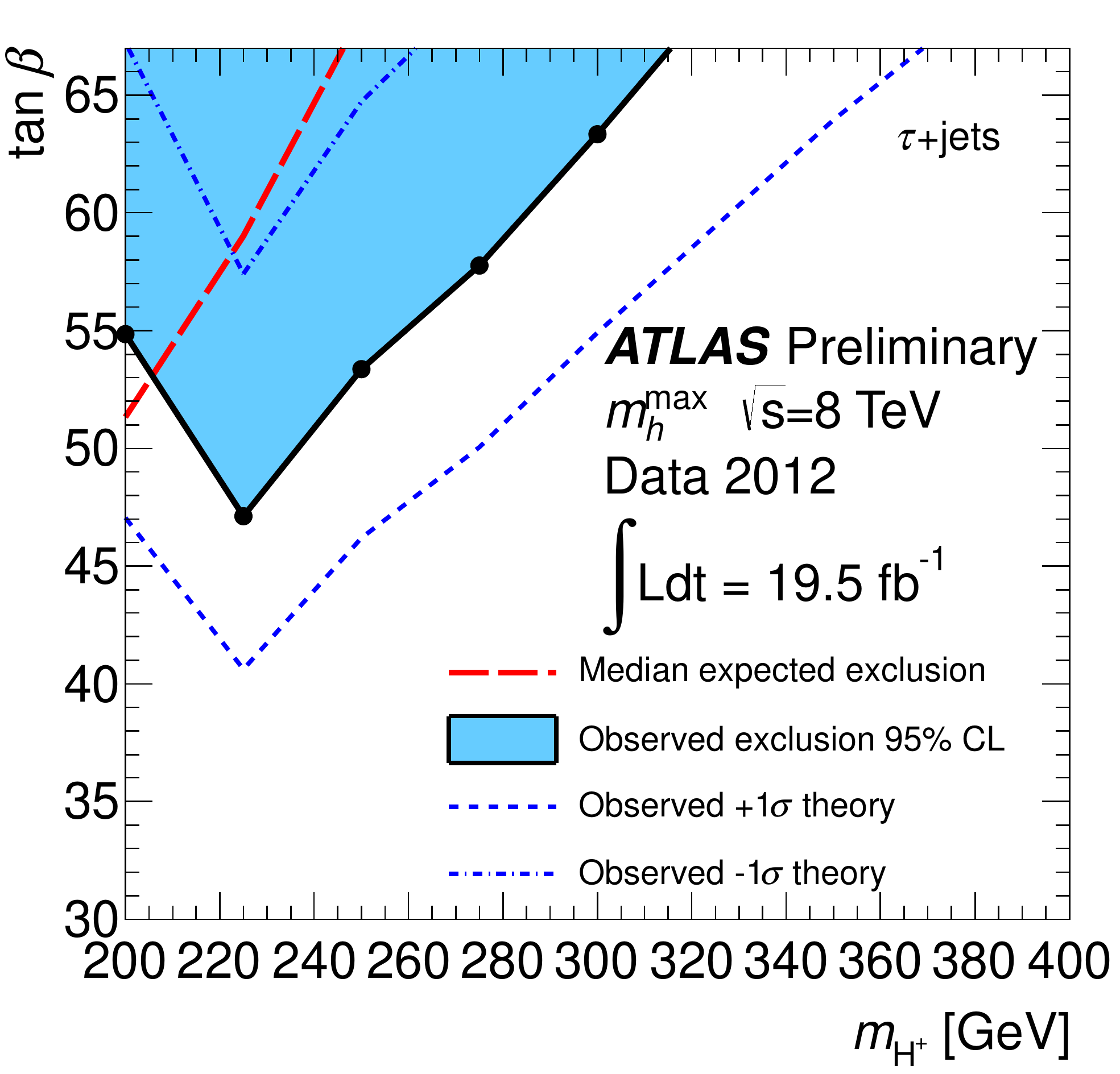}}
\vspace*{8pt}
\caption{The left-hand plots show the data and background predictions
  for the ATLAS $H^+$ boson search\protect\cite{ATLAS_HPlus} as function of $m_T$
, both for the light (top) and heavy
  (bottom) Higgs mass selections. Expected Higgs signals for Higgs
  masses of 130 and 250~GeV are also superimposed, scaled up by a
  factor of ten, assuming $BR(t \rightarrow b H^+) = 0.9\%$ and $\tan
  \beta = 50$, respectively. The right-hand plots show the resulting
  model-dependent limits in the $m_{H^+}$ and $\tan \beta$ parameter
  plane, again for the light (top) and heavy
  (bottom) Higgs mass selections\protect\cite{ATLAS_HPlus}.}
\label{Fig:ATLASHPlus}
\end{figure}
Discovery of a charged Higgs boson would be an immediate indication of
physics beyond the SM. For $\tan \beta>3$, the dominant
decay mode for light charged Higgs bosons is $H^+ \rightarrow \tau
\nu_{\tau}$, for heavy charged Higgs bosons the branching fraction to $\tau
\nu_{\tau}$ can still be sizable. The main production modes depend on the mass of the
charged Higgs. For $m(H^+)<m_t$, it can be produced in top quark
decays, while for $m(H^+)>m_t$, associated production together with
top quarks takes over. The ATLAS analysis shown here\cite{ATLAS_HPlus}
uses $t\bar t$ events with a $\tau$ lepton decaying hadronically in the final state, 
with veto on any other
leptons. The analysis requires at least three or four jets, with at least one of
them b-tagged, and large missing $E_T$. The discriminating variable is
the transverse invariant mass of the $\tau$ products combined with the
missing $E_T$, defined as 
\begin{equation}
m_T = \sqrt{ 2p_T^\tau E_T^{miss}(1 - \cos \Delta \phi_{\tau,miss}) }
\label{atlasMT}
\end{equation}
where $\Delta \phi_{\tau,miss}$ is the azimuthal angle between the
hadronic decay products of the $\tau$ lepton and the direction of the
missing transverse momentum. Main backgrounds are general $t\bar t$,
single top, W/Z+jets and di-boson production, as well as QCD.

The full 2012 dataset is used for this measurement. The low and
high mass selections cover $H^+$ mass ranges of 90--160 and
180--600~GeV, respectively. Figure~\ref{Fig:ATLASHPlus} shows the
$m_T$ spectra as well as the model dependent limits for both
selections. No evidence for a $H^+$ signal is found. At low masses,
large parts of the MSSM parameter space are excluded, while at high
masses relatively large values of $\tan \beta$ are still allowed.

\section{Generic 2HDM Searches}

\begin{figure}[htbp]
\centerline{\includegraphics[height=5.0cm]{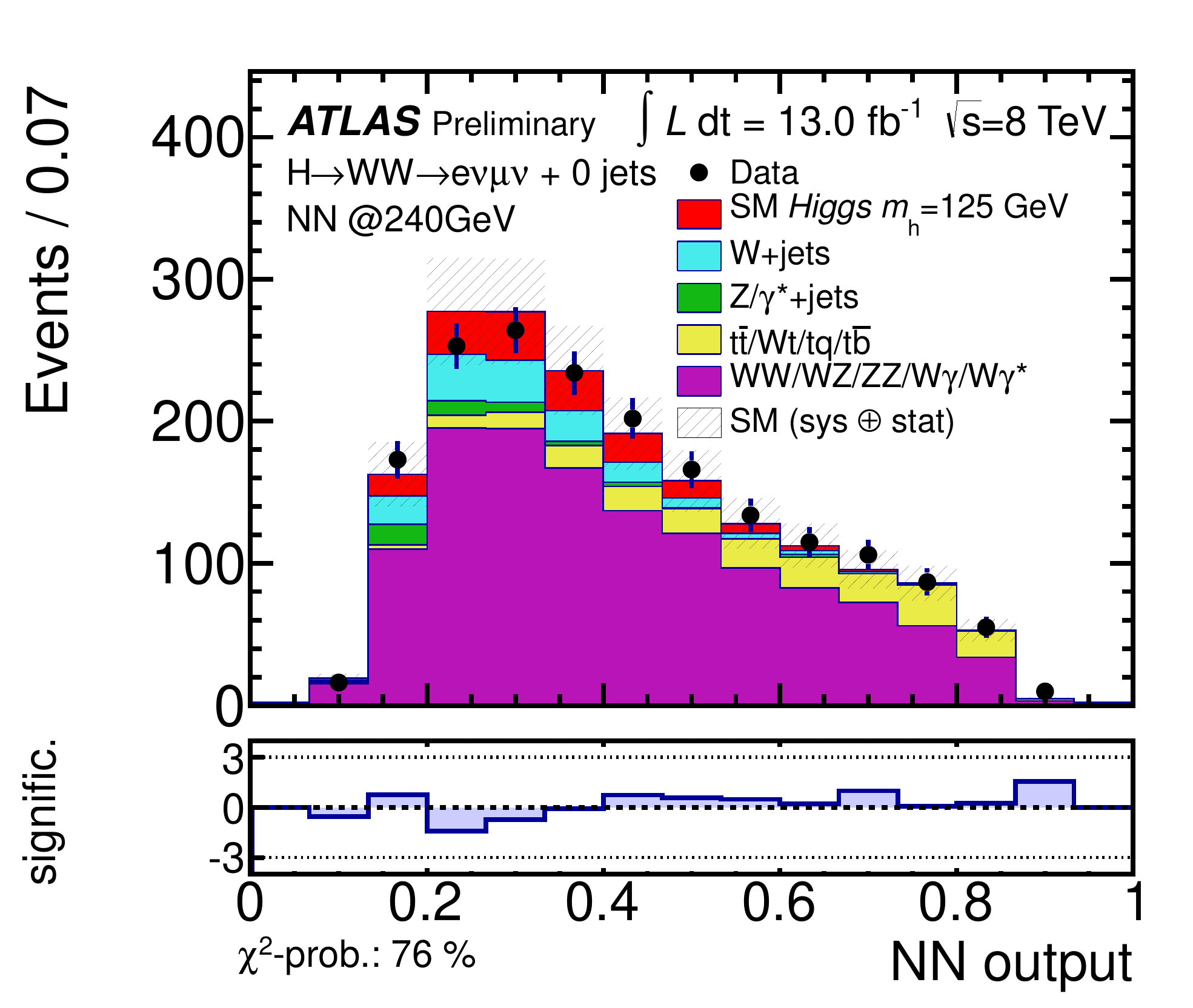}
\includegraphics[height=5.0cm]{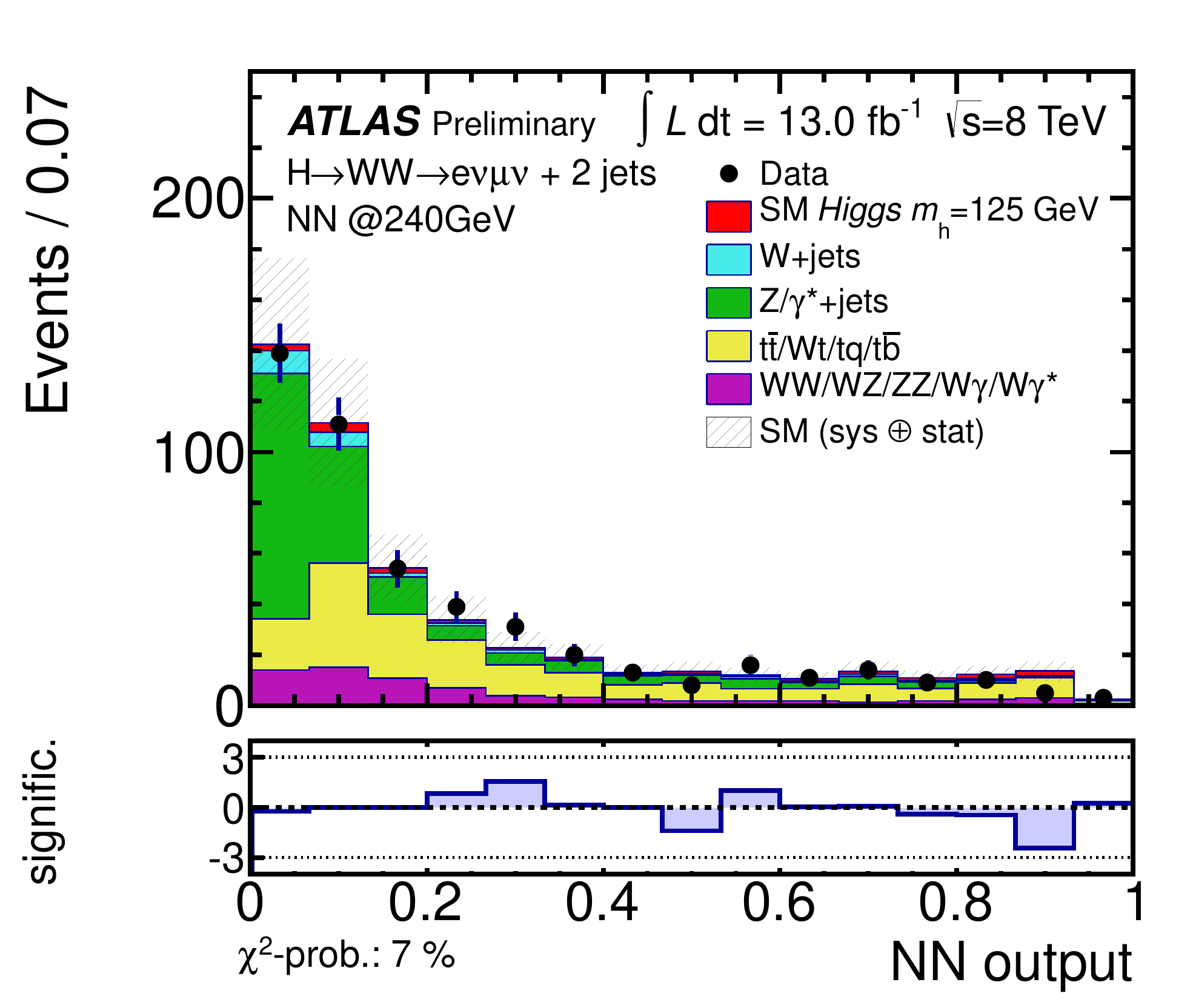}}
\vspace*{8pt}
\caption{Distributions of the neural network discriminant variable for
  the ATLAS $h / H \rightarrow WW^{(*)} \rightarrow e\nu\mu\nu$
  analysis\protect\cite{ATLAS_2HDM}, optimized for a Higgs mass of
  240~GeV, for the GGF (left) and VBF selections (right).}
\label{Fig:ATLAS2HDMDiscriminant}
\end{figure}

While the MSSM is a very specific implementation of a Higgs sector
motivated by supersymmetry, the Two-Higgs Doublet Model (2HDM) is a
phenomenological approach, which allows data interpretations based on
two Higgs doublets without committing to a particular theory. As
opposed to MSSM at tree level, the 2HDM can accomodate CP violation,
as well as flavor-changing couplings. Examples for categories of 2HDM models with
natural flavor conservation are the Type-I, in which all quarks
couple only to one Higgs doublet, and the Type-II, in which the
up-type quarks couple to one and the down-type quarks couple to the
other Higgs doublet. Key parameters are $\tan \beta$, which is the
ratio of the vacuum expectation values of the two Higgs doublets, and
the scalar mixing angle $\alpha$. The MSSM is a special case of a
Type-II 2HDM. As the MSSM, the 2HDM feature three neutral and two charged Higgs bosons.

\begin{figure}[htbp]
\centerline{\includegraphics[height=5.0cm]{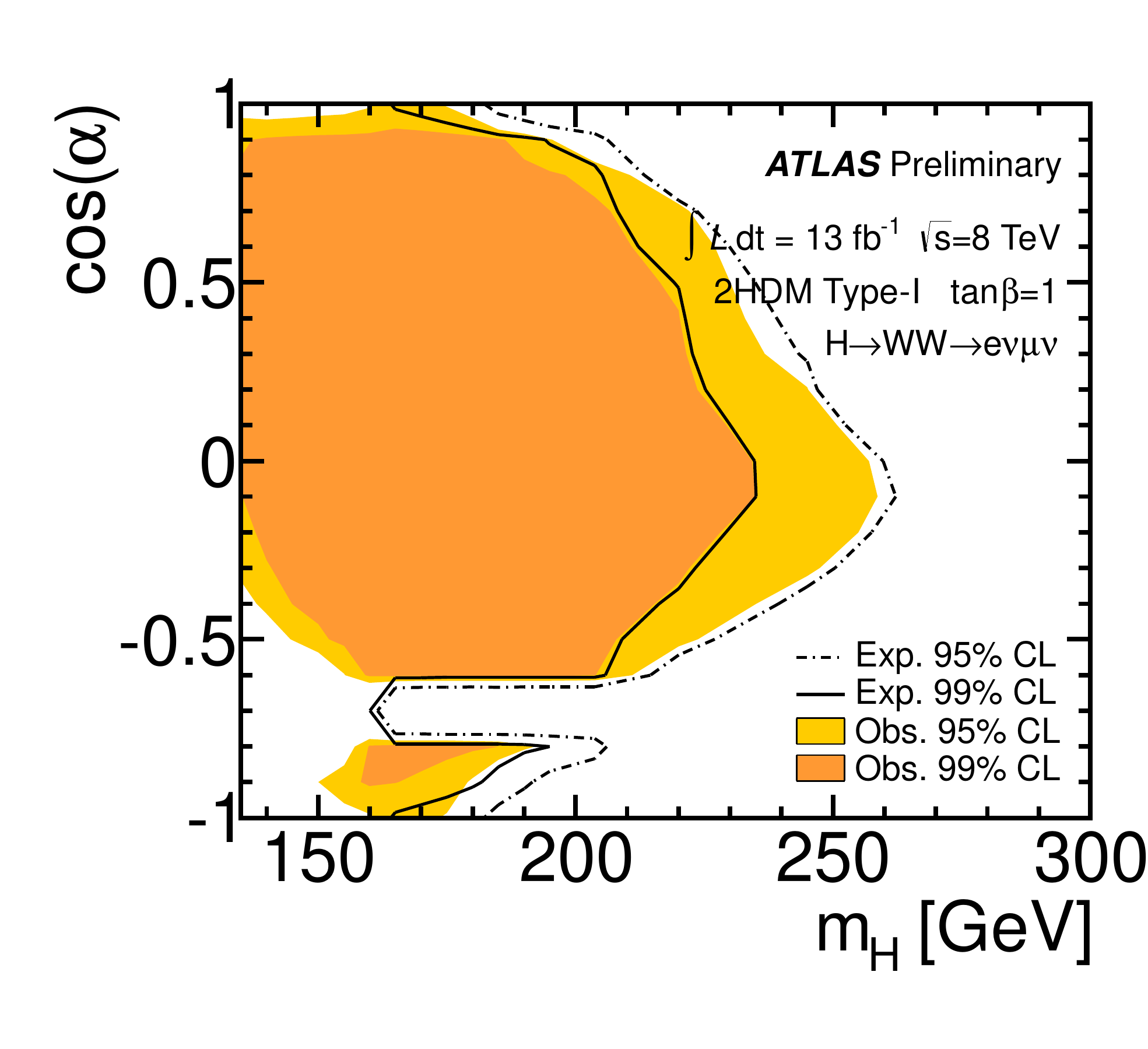}
\includegraphics[height=5.0cm]{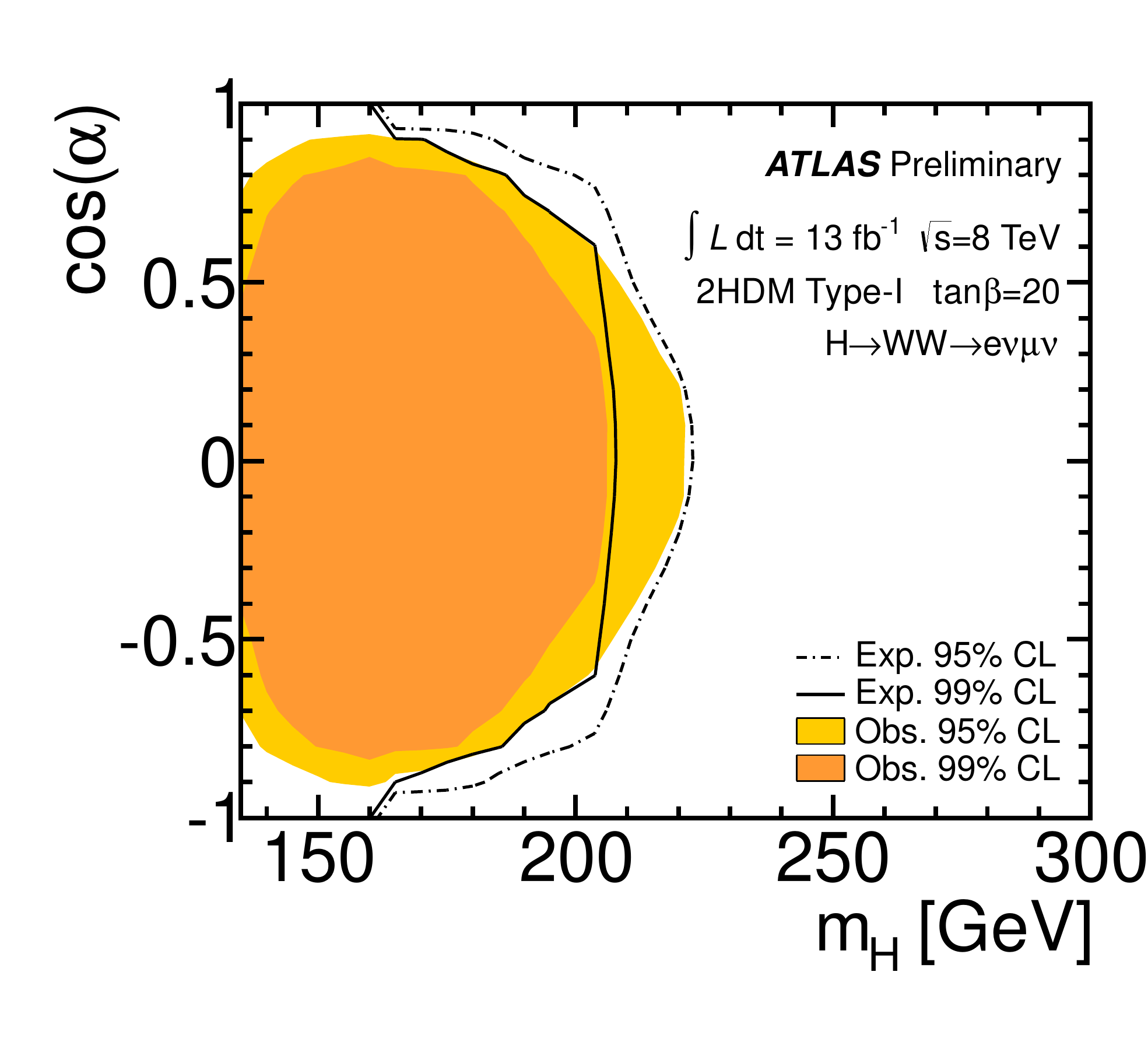}}
\centerline{\includegraphics[height=5.0cm]{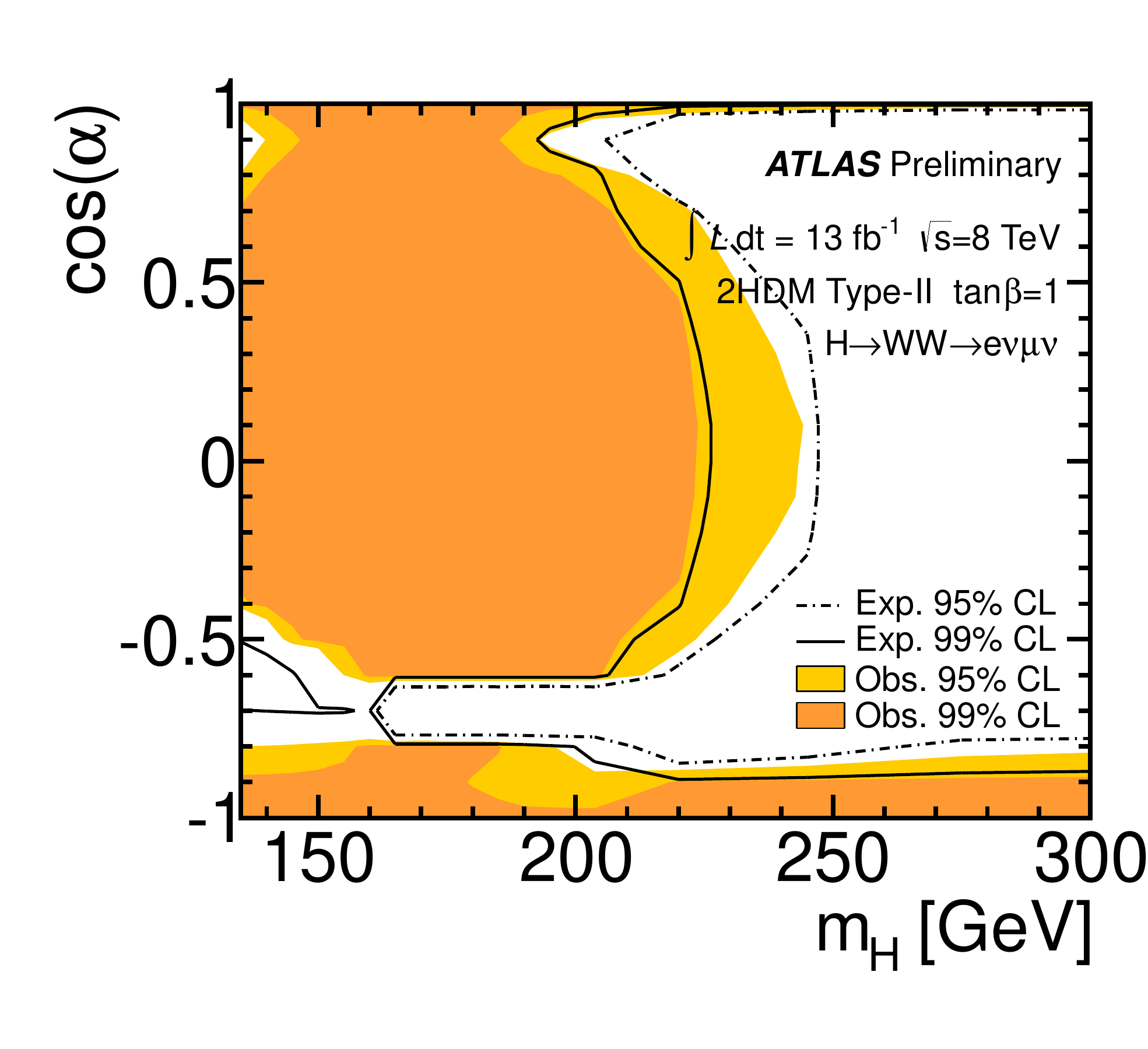}
\includegraphics[height=5.0cm]{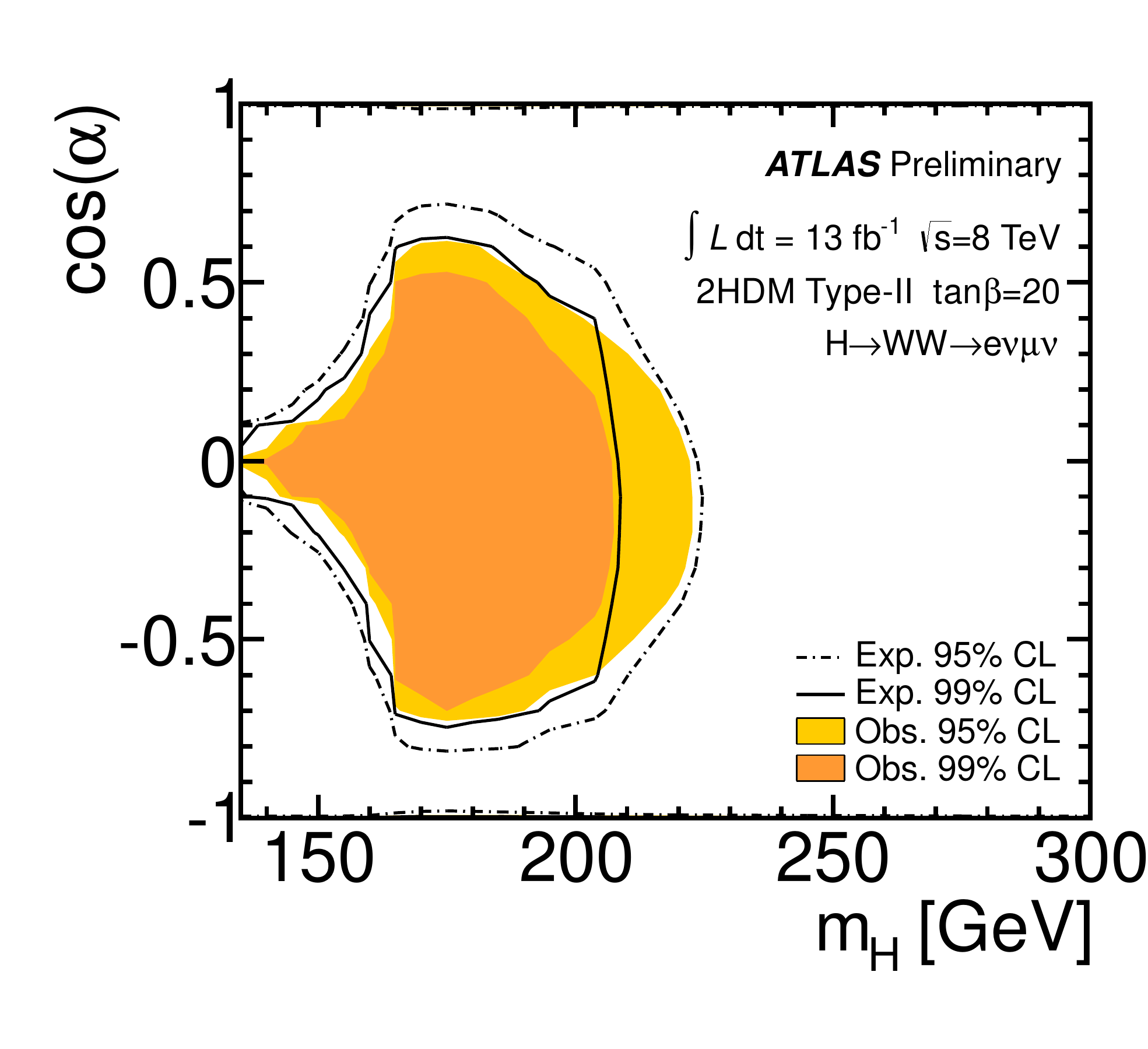}}
\vspace*{8pt}
\caption{Exclusion contours in the $\cos \alpha - m_H$ parameter space
  for the ATLAS $h / H \rightarrow WW^{(*)} \rightarrow e\nu\mu\nu$
  analysis\protect\cite{ATLAS_2HDM} . The top row shows Type-I and the bottom row Type-II 2HDM
  models. The left and right column plots assume $\tan \beta=1$ and
  $\tan \beta=20$, respectively.}
\label{Fig:ATLAS2HDMContours}
\end{figure}
A recent analysis from ATLAS\cite{ATLAS_2HDM} searches for decays of
the scalar Higgs bosons h and H in the channel $h / H \rightarrow
WW^{(*)} \rightarrow e\nu\mu\nu$, assuming $m_h=125$~GeV. The
pseudo-scalar A does not decay into W pairs. The analysis requires
exactly two leptons of opposite charge and 
missing energy. In order to cover the 
GGF and vector boson fusion (VBF) production mechanisms, either
 zero or two jets are required, respectively. A neural network combining several kinematic variables is
trained for each mass point to enhance the signal over the background.
The Higgs boson observed near 125~GeV is also treated as background. The
resulting distributions of the neural network discriminant are shown
in Figure~\ref{Fig:ATLAS2HDMDiscriminant} for an assumed Higgs mass of
240~GeV. The data are compared to a model of simulated events representing the various background processes.
No indication of a
signal is seen. The results are translated into exclusion
contours within the 2HDM parameter space. Figure~\ref{Fig:ATLAS2HDMContours} shows
examples for such contours for two values of $\tan \beta$, each for
Type-I and Type-II 2HDM models. For low masses, significant parts of
the $\cos \alpha$ range are excluded.

\section{NMSSM Higgs Search}
\begin{figure}[htbp]
\centerline{\includegraphics[height=6.0cm]{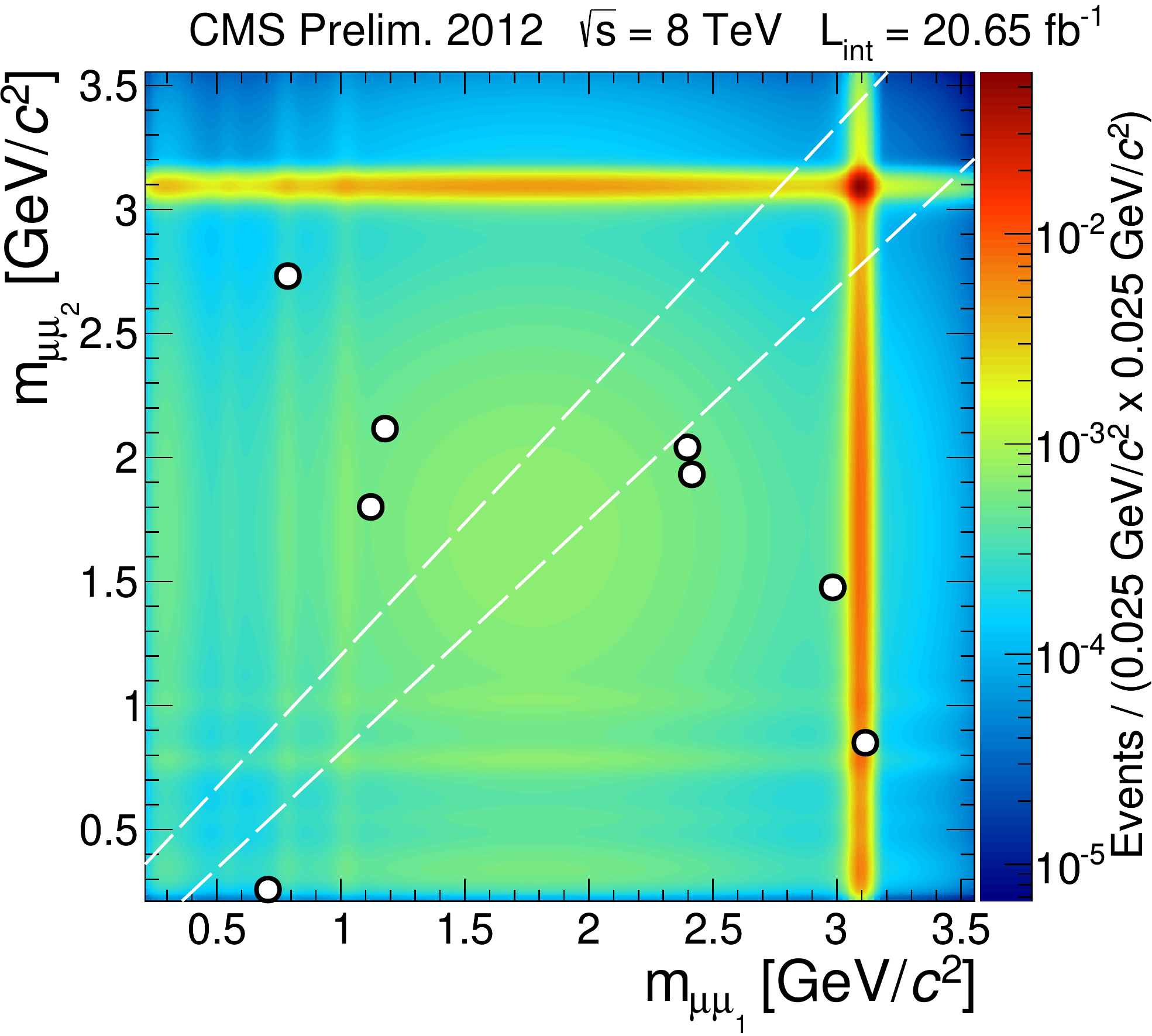}
\includegraphics[height=6.0cm]{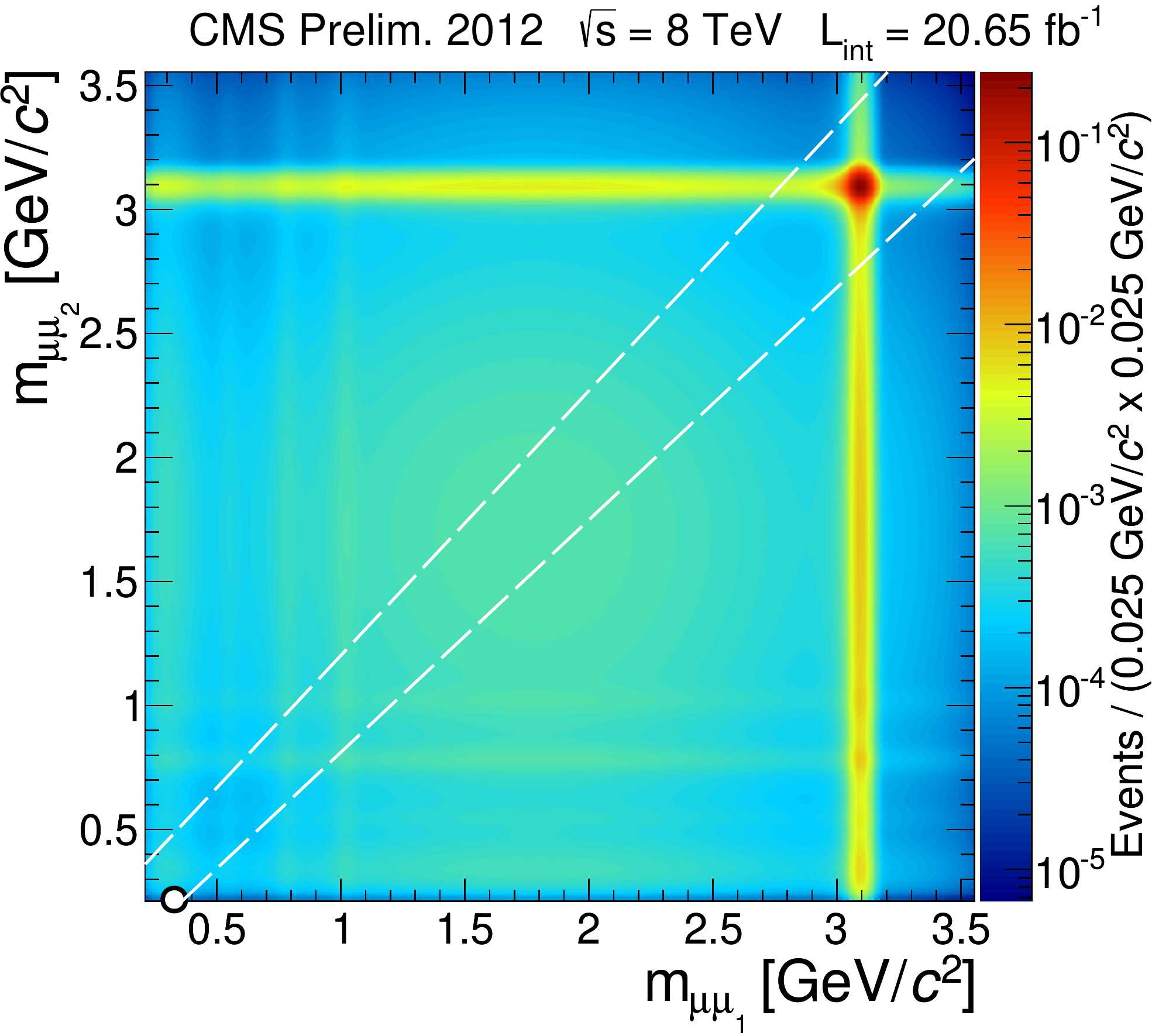}}
\vspace*{8pt}
\caption{Distribution in the invariant masses of the two muon pairs,
  in the off-diagonal control region (left) and the diagonal signal
  region (right) in the CMS NMSSM search\protect\cite{CMS_4mu}. The empty circles show the surviving events in the
  data. The histogram represented by shades shows the $b\bar b$
  background template.}
\label{Fig:CMS_NMSSM_dimass}
\end{figure}
The next-to-minimal supersymmetric model (NMSSM) features two complex
Higgs doublets and an additional scalar field. The physical states are
mixtures: three CP-even ($h_1$, $h_2$, $h_3$), two CP-odd
($a_1$, $a_2$), and two charged bosons ($h^\pm$). The NMSSM requires
less fine tuning for the Higgs mass, and solves the so-called ``$\mu$
problem'' of the MSSM.

A recent CMS analysis\cite{CMS_4mu} searches for the decay of a non-standard Higgs
into two very light bosons, resulting in two boosted pairs of
muons. The corresponding NMSSM interpretation is a decay chain
$h_{1,2} \rightarrow a_1 a_1 \rightarrow (\mu\mu) (\mu\mu)$, where
either the $h_1$ or the $h_2$ could correspond to the boson observed
near 125~GeV, and $a_1$ is a new CP-odd Higgs boson lighter than
twice the $\tau$ mass. But also an interpretation within dark-SUSY
models is possible, based on the decay chain $h \rightarrow 2 n_1
\rightarrow 2 n_D + 2 \gamma_D \rightarrow 2 n_D +  (\mu\mu)
(\mu\mu)$, where $n_1$ is the lightest visible neutralino, $n_D$ is a
light dark fermion and $\gamma_D$ a light massive dark photon with
weak couplings to SM particles.

\begin{figure}[htbp]
\centerline{\includegraphics[height=6.0cm]{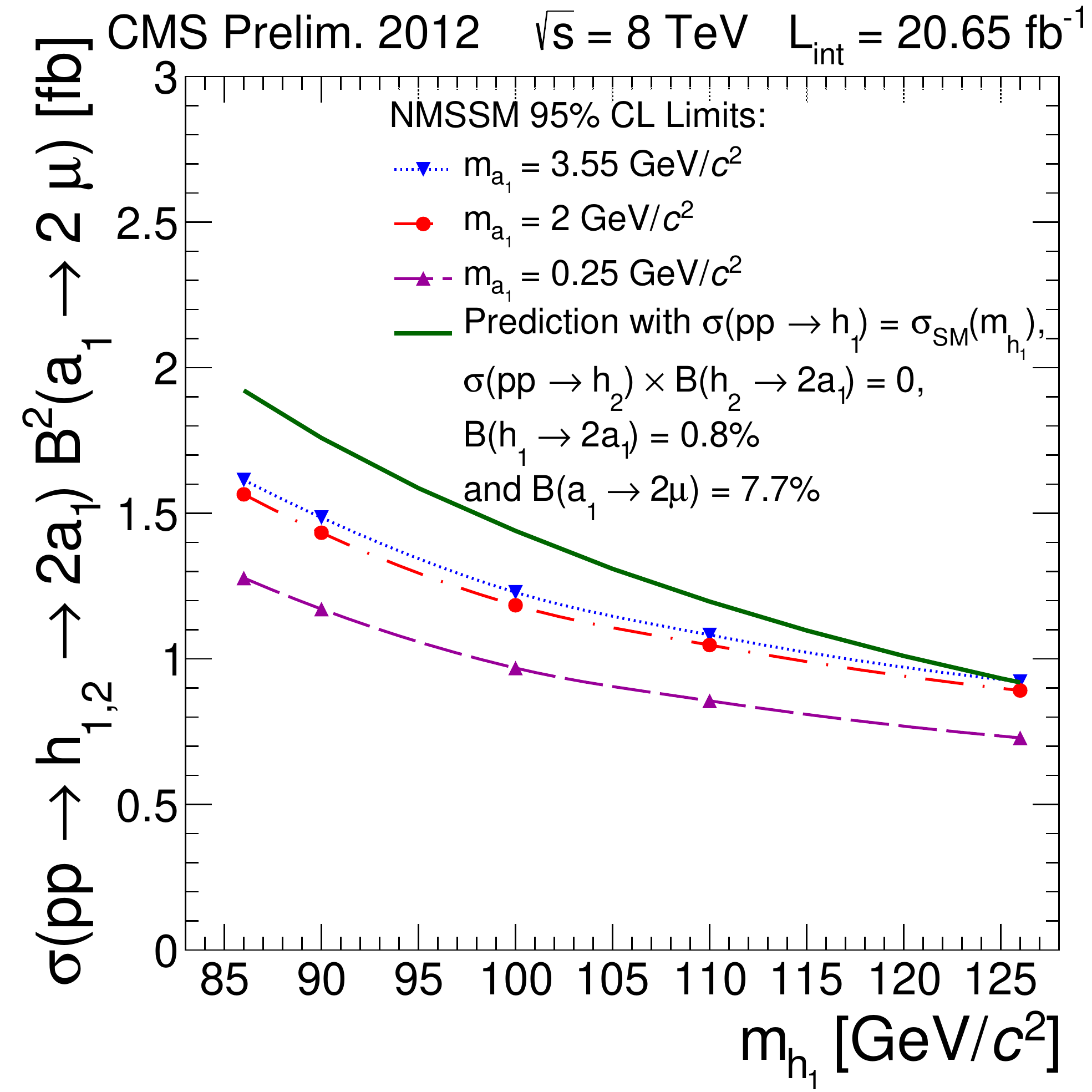}
\includegraphics[height=6.0cm]{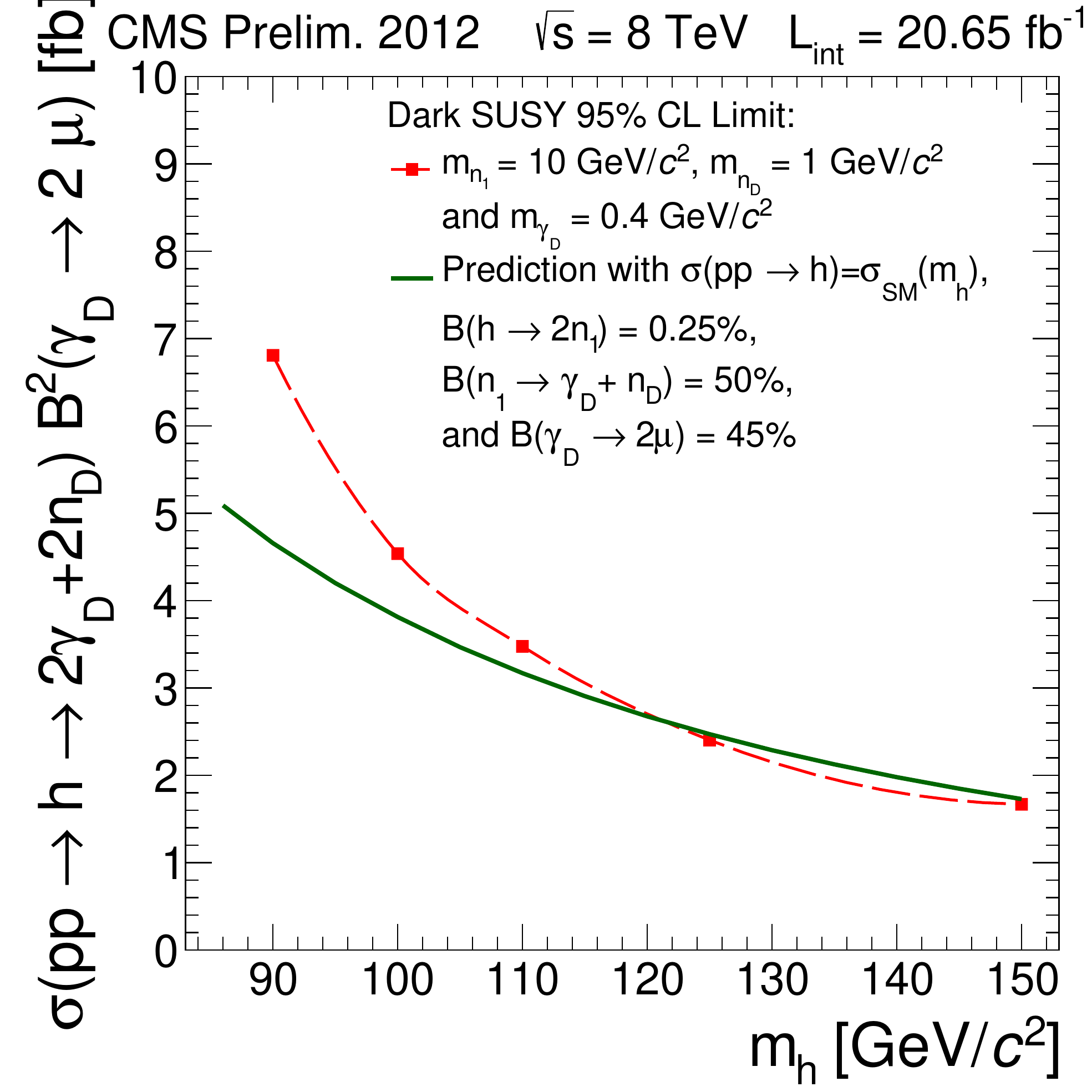}}
\vspace*{8pt}
\caption{Interpretation of the CMS results in the NMSSM (left) and the dark-SUSY
  (right) benchmark models. (See text for details).}
\label{Fig:CMS_NMSSM_UL}
\end{figure}
The analysis selects events with two isolated, boosted muon pairs,
considering the mass ranges $0.25<m_a<3.55$~GeV and $m_h>86$~GeV. The
signal region is defined by requiring the masses of the two muon pairs
to be equal. Main backgrounds in the analysis are direct production of
$J/\Psi$ pairs, and $b\bar b$ production with subsequent di-muon
decays, either semi-leptonic or via quarkonia resonances. The
$b\bar b$ background is estimated from a bb-enriched control sample and
the double-$J/\Psi$ production from
PYTHIA. Figure~\ref{Fig:CMS_NMSSM_dimass}~(left) shows the distribution in
the space of the two di-muon masses with the diagonal signal region
still blinded. Eight events are observed in the off-diagonal
sideband. After unblinding, only one event is observed in the diagonal
signal region (Figure~\ref{Fig:CMS_NMSSM_dimass}~(right)), consistent with an
expected background of $3.8 \pm 2.1$ events. The results are
interpreted in the context of NMSSM and dark-SUSY
models. The NMSSM interpretation (Figure~\ref{Fig:CMS_NMSSM_UL}~(left)) shows 95\% CL upper 
limits on the cross section times branching fraction vs. the Higgs mass 
$m_{h_1}$ for different values of $m_{a_1}$. The dark-SUSY
interpretation (Figure~\ref{Fig:CMS_NMSSM_UL}~(right)) displays 
corresponding limits vs. $m_h$ for a simplified scenario for 
$m_{n_1}=10$~GeV and $m_{n_D}=1$~GeV and
$m_{\gamma_D}=0.4$~GeV. The comparison shows that the experimental
limits of this analysis are already able to exclude certain models.

\section{Searches for Invisible Higgs Bosons}
If a Higgs boson would decay with a significant probability into
final states consisting entirely of invisible particles, this might
still be measurable if the Higgs is produced in association with other,
well detectable particles. Searches of invisible Higgs decays in
association with Z bosons have lead to upper limits on the invisible
branching fraction of 65\%
(ATLAS\cite{ATLASHInv}) and 75\% (CMS\cite{CMSHInv}) for a
  Higgs with SM cross section at a mass of 125~GeV at 95\% confidence level, leaving still
  plenty of room for such decays.

\begin{figure}[htbp]
\centerline{\includegraphics[height=6.0cm]{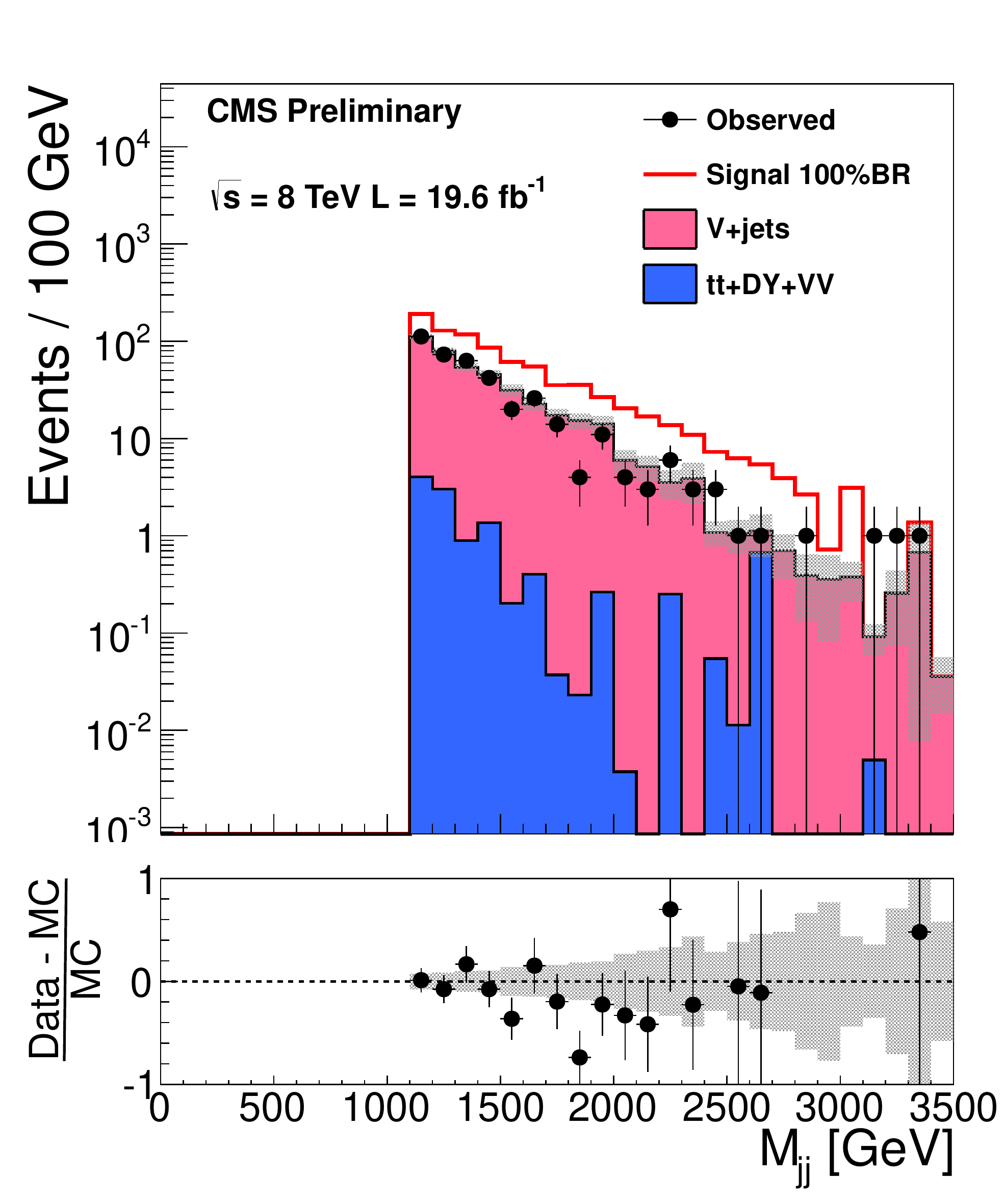}
\includegraphics[height=5.0cm]{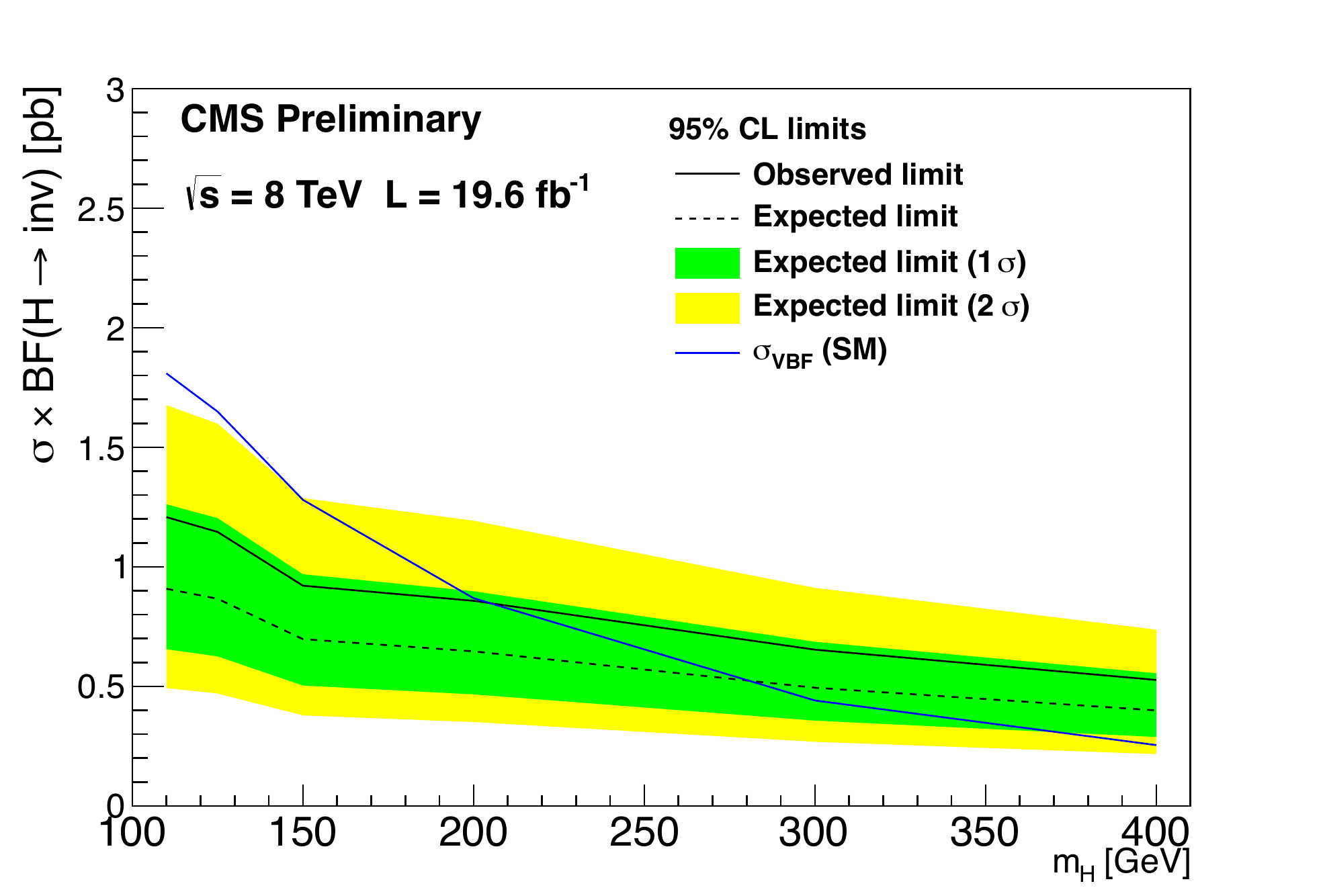}}
\vspace*{8pt}
\caption{Left: Di-jet mass distribution for VBF events together with
  the estimated backgrounds in the invisible Higgs search of CMS\protect\cite{CMSHInvVBF}. The
  signal expected for a 100~\% invisible branching fraction is also
  shown. Right: Upper limits for $\sigma \times BR$ into invisible
  final states. The full VBF Higgs production cross section is also shown.}
\label{Fig:CMSHInvVBF}
\end{figure}
  Recently, invisible Higgs decays have also been searched for by CMS in VBF
  topology\cite{CMSHInvVBF}, which has a higher cross section than
  associated Z production. The final state is characterized by two
  scattered jets with a large rapidity gap, very little other activity
  in the event and large missing $E_T$. Special triggers combining VBF and missing $E_T$ signatures are
  crucial for this analysis, and large efforts have been undertaken to
  reduce the QCD background. The signal is searched for in the
  invariant mass of the two jets. The signature of an invisible
  Higgs should manifest as an excess which is growing with the dijet
  mass. Figure~\ref{Fig:CMSHInvVBF} (left) shows the observed
  distribution in the data with the background expectation and the
  expected signal in case the Higgs near 125~GeV would decay to 100~\%
  into invisible modes. No signal is observed.
  Figure~\ref{Fig:CMSHInvVBF} (right) shows the obtained upper limits
  on the cross section times branching fraction
  as a function of the Higgs mass, in comparison with the full VBF cross
  section. At $m_H=125$~GeV, the upper limit on the invisible
  branching fraction is 69~\%.

\section{Summary}
The observation of SM-like properties of the established Higgs state
near 125~GeV does not imply that the Higgs sector must have SM
structure. The best way of clarification is the direct search for
additional Higgs signatures.

A broad attack is launched to clarify whether the Higgs sector reaches
beyond the SM. In the context of MSSM, at low masses $m_A$ the limits
from LEP and LHC start to close. Large values of $m_A$ and $\tan
\beta$ are still possible. The constraints from the $H^+$ searches have
significantly improved. Recently, analyses interpret their findings
also in the 2HDM approach. Concerning the NMSSM, only few channels
have been targeted so far, and there is still a wide open range of
possibilities. The search for invisible Higgs particles and decay
modes yields first results in vector boson associated production and
VBF signatures, but the limits on the invisible
branching fraction are still large.

In summary, the Higgs searches beyond the SM have just scratched the
surface. Many LHC analyses are being updated with the full 8~TeV
statistics, and the Run-II of the LHC at $\approx$13~TeV will further
extend the reach towards higher masses.





\end{document}